\documentclass[traditabstract]{aa}
\usepackage{graphicx,psfig,color}
\usepackage{multirow}

\newcommand{\degree} {$^{\rm o}$}

\newcommand{\simless}{\mathbin{\lower 3pt\hbox
      {$\rlap{\raise 5pt\hbox{$\char'074$}}\mathchar"7218$}}} %< or of order
\newcommand{\simgreat}{\mathbin{\lower 3pt\hbox
     {$\rlap{\raise 5pt\hbox{$\char'076$}}\mathchar"7218$}}} %> or of order

\begin{document}

\newcommand{\hi}{\ion{H}{i}~}
\newcommand{\hii}{\ion{H}{ii}~}

   \title{The effect of local optically thick regions in the long-wave emission of young circumstellar disks}

   \author{L. Ricci         \inst{1,2}
\and       F. Trotta        \inst{1,3} 
\and       L. Testi         \inst{1,3} 
\and       A. Natta         \inst{3,4} 
\and       A. Isella        \inst{2} 
\and       D. J. Wilner        \inst{5} 
}

%  \offprints{O. Tiret }

   \institute{  European Southern Observatory,
   Karl-Schwarzschild-Strasse 2, D-85748 Garching, Germany
            \and
                     Division of Physics, Mathematics and Astronomy, California Institute of Technology, MC 249-17, Pasadena, CA 91125, USA                
            \and
                     INAF - Osservatorio Astrofisico di Arcetri, Largo Fermi 5, I-50125 Firenze, Italy
            \and
                     School of Cosmic Physics, Dublin Institute for Advanced Studies, 31 Fitzwilliam Place, Dublin 2, Ireland     
            \and
Harvard-Smithsonian Center for Astrophysics, 60 Garden Street, Cambridge, MA 02138, USA
                     }

   \date{Received XXX 2010/ Accepted YYY ZZZZ}

   \titlerunning{Optically thick regions in protoplanetary disks}

   \authorrunning{Ricci et al.}

\abstract{ Multi-wavelength observations of protoplanetary disks in the sub-millimeter continuum have measured spectral indices values which are significantly lower than what is found in the diffuse interstellar medium. Under the assumption that mm-wave emission of disks is mostly optically thin, these data have been generally interpreted as evidence for the presence of mm/cm-sized pebbles in the disk outer regions. 

In this work we investigate the effect of possible local optically thick regions on the mm-wave emission of protoplanetary disks without mm/cm-sized grains. A significant local increase of the optical depth in the disk can be caused by the concentration of solid particles, as predicted to result from a variety of proposed physical mechanisms. We calculate the filling factors and implied
overdensities these optically thick regions would need to significantly
affect the millimeter fluxes of disks, and we discuss their plausibility.

We find that optically thick regions characterized by relatively small filling factors can reproduce the mm-data of young disks without requesting emission from mm/cm-sized pebbles. However, these optically thick regions require dust overdensities much larger than what predicted by any of the physical processes proposed in the literature to drive the concentration of solids. 
We find that only for the most massive disks it is possible and plausible to imagine that the presence of optically thick regions in the disk is responsible for the low measured values of the mm spectral index. For the majority of the disk population, optically thin emission from a population of large mm-sized grains remains the most plausible explanation. The results of this analysis further strengthen the scenario for which the measured low spectral indices of protoplanetary disks at mm wavelengths are due to the presence of large mm/cm-sized pebbles in the disk outer regions.       
}

%{$ $}{$ $}{$ $}{$ $}


\maketitle

%---------------------------------------------------------------

\section{Introduction}

Planets around solar-like stars are thought to originate from the material contained in a circumstellar ``protoplanetary'' disk. 
Observations of protoplanetary disks around pre-main sequence (PMS) stars at optical and infrared wavelengths have provided evidence for the presence of dust grains as large as at least a few $\mu$m in many of these systems. Since these grains are larger than the submicron-sized grains found in the interstellar medium (ISM, e.g. Mathis et al.~\cite{Mat77}), these observational results have been interpreted in terms of dust grain growth from an original ISM-like dust population in the disk. These are the first steps of growth of solid particles which may potentially lead to the formation of planetesimals and then planetary bodies. 

In order to investigate the presence of larger grains in the disk, observations at longer wavelengths are needed. Furthermore, since the dust opacity decreases at longer wavelengths, whereas infrared observations are sensitive to emission from the disk surface layers, observations in the millimeter probe the denser disk midplane, where the whole process of planetesimal formation is supposed to occur. 

In the last two decades several authors measured relatively shallow slopes $\alpha$ of the Spectral Energy Distribution (SED; $F_{\nu} \sim \nu^{\alpha}$ with $\alpha \sim 2-3$) of protoplanetary disks at sub-mm and mm wavelengths (e.g. Beckwith \& Sargent~\cite{Bec91}, Wilner et al.~\cite{Wil00}, Testi et al.~\cite{Tes01,Tes03}, Natta et al.~\cite{Nat04}, Wilner et al.~\cite{Wil05}, Andrews et al.~\cite{And05}, Rodmann et al.~\cite{Rod06}, Ricci et al.~\cite{Ric10a}).
Under the assumption of completely optically thin emission and if the emitting dust is warm enough to make the Rayleigh-Jeans approximation hold true at these wavelengths, the SED spectral index $\alpha$ is directly linked to the spectral index $\beta$ of the dust opacity coefficient $\kappa_{\nu}$\footnote{At these long wavelengths $\kappa_{\nu}$ is well approximated by a power-law.} through $\beta=\alpha-2$.    
In this way, the measured low values of $\alpha$ translate into values of $\beta \simless 1$ which are significantly lower than the value of $1.5-2$ associated to the ISM (Mathis et al.~\cite{Mat77}). For all the reasonable models of dust analyzed so far, the obtained values of $\beta$ for young disks can be interpreted only if grains have grown to sizes of at least a few millimeters in the outer disk regions (see e.g. Draine~\cite{Dra06}, Natta et al.~\cite{Nat07}).

The presence of mm/cm-sized solid particles in the outer disk poses serious problems to the models of dust evolution in young circumstellar disks. Because of pressure, gas moves with sub-keplerian velocity, and thus it is slower than the solid component which is rotating at keplerian velocity. This generates a gas headwind felt by the solid particles, causing them to lose angular momentum and spiral inward in the disk. Dynamical models of solids in protoplanetary disks show that mm/cm-sized particles in the outer regions of disks are expected to rapidly drift toward the inner disk (Weidenschilling~\cite{Wei77}). As a consequence, these particles should not be present in the disk outer regions. This expectation is challenged by the observations of mm-slope of protoplanetary disks, if interpreted as due to large grains emission. Furthermore, Birnstiel et al. (\cite{Bir10a}) have recently shown that even if radial drift is completely halted by some physical mechanisms, the low mm-spectral indices ($\approx 2-2.5$) measured for the faintest disks cannot be reproduced by models of dust evolution in disks with low masses. This is because for these disks the gas densities in the outer regions are so low that mm/cm grains should be never formed there through sticking after mutual grain collision. 

Another \textit{a priori} possible scenario for the interpretation of the measured low values of the mm-spectral indices is that a significant fraction of emission at these wavelengths come from optically thick regions in the disk. In this case, the spectral index of the SED would reflect only the spectral index of the Planck function, which is 2 for emission in the Rayleigh-Jeans regime. This value is consistent with that measured for a large population of young circumstellar disks (e.g. Rodmann et al.~\cite{Rod06}, Ricci et al.~\cite{Ric10a}). 

Under the assumption of smooth disks, the innermost regions of typical T Tauri disks ($R \simless 10-20$~AU) can be dense enough to be optically thick even at long mm-wavelengths (e.g. Beckwith et al.~\cite{Bec90}, Testi et al.~\cite{Tes01}). However, interferometric observations of disks can constrain the physical structure of disks and quantify the impact of this optical-depth effect. The general conclusion of these studies is that disks are typically found to be relatively large, with outer radii of the order of $\sim 100$~AU or more. By modelling the mm-wave emission one finds that the typical contribution of the optically thick innermost regions to the total fluxes is nearly insignificant at wavelengths $\simgreat~1$~mm (see e.g. Testi et al.~\cite{Tes03}, Wilner et al.~\cite{Wil05}, Isella et al.~\cite{Ise09}, Guilloteau et al.~\cite{Gui11}).  

In the last years several different physical processes with the potential of concentrating particles in disks have been proposed as possible triggering mechanisms for the formation of planetesimals (see Chiang \& Youdin~\cite{Chi10}). These all lead to a local increase of the particle density in some regions of the disk. If the density gets high enough, these regions might become optically thick even at mm wavelengths. Where in the disk these particle concentrations are supposed to occur depends on the particular physical mechanism driving the process, but they are not necessarily expected only in the innermost disk regions.

If this scenario was viable, the spectral index of the SED would not carry the information on the grain sizes (through $\beta$) and no constraints on that property could be derived from the observations. 
Therefore it is important to investigate the potential effect of optically thick disk regions on the disk total emission at millimeter wavelengths to understand whether the measured fluxes can be really used to test models of dust evolution in protoplanetary disks.  
With this work we want to answer the following questions: 1) could the observed low values of the (sub-)millimeter spectral indices of disks be explained by local concentrations of small particles (sizes $<$~1 mm) \textit{rather than} by the widespread presence of mm/cm-sized pebbles? 2) If yes, which characteristics do they need to have? 3) Are the required local concentrations of small particles in disks physically plausible?  

In Section~\ref{sec:sample} we present the sample of protoplanetary disks that we will use for our analysis. Section~\ref{sec:analysis} outlines the disk and dust models considered for the analysis. The results are described in Section~\ref{sec:results}, whereas a discussion on the impact of the optically thick regions on high-angular resolution observations of disks in the sub-millimeter, and on the physical plausibility of the required structures are presented in Section~\ref{sec:discussion}. The main conclusions of this work are summarized in Section~\ref{sec:summary}.

\section{Sample}
\label{sec:sample}

We describe here the sample of disks that we will consider in the following analysis. 
This is made of the samples of low-mass young stellar objects (YSOs) in the Taurus, Ophiuchus and Orion Nebula SFRs by Ricci et al.~(\cite{Ric10a},~\cite{Ric10b},~\cite{Ric11a},~\cite{Ric11b}, respectively). These sources have been selected (1) by being low-mass Class II YSOs with no evidence of extended emission from a parental envelope, (2) by having known sub-mm/mm SED, through at least one detection in the 0.45 $\leq \lambda \leq$ 1.3 mm spectral range and a detection at $\lambda \approx 3$~mm, 
%(3) \textbf{by having well characterized stellar properties through optical-NIR spectroscopic/photometric data}, 
(3) as well as no evidence of any stellar companion at spatial separations of about $10-400$~AU that would likely affect the structure of the disk outer regions, to which observations at sub-mm/mm wavelengths are most sensitive to. 

In addition to these 47 sources, we consider here also the disks around BP Tau, DQ Tau, V836 Tau for which we obtained new CARMA observations and which satisfy the selection criteria described above, as detailed in Appendix A. 

For a detailed discussion on the completeness of the selected samples we refer to the Ricci et al.~(\cite{Ric10a},~\cite{Ric10b},~\cite{Ric11a},~\cite{Ric11b}) papers. Here, we note that the second selection criterion listed above implies that the selected disks are relatively bright in the sub-mm and therefore relatively massive. A list of the disk mass constrained from the sub-mm/mm SED of each disk is reported in Appendix B. As can been seen from Table~6, disks with masses as low as about $10^{-5}~M_{\odot}$ in dust, or $\sim 10^{-3}~M_{\odot}$ in gas (given a dust-to-gas mass ratio of 0.01), are included in our sample.

\section{Analysis}
\label{sec:analysis}

In this Section we present the method adopted to explore the impact of optically thick regions in the total emission of a disk at sub-mm/mm wavelengths. Section~\ref{sec:disk_structure} describes the disk model used in our analysis and how the optically thick regions are included in the disk structure.
Section~\ref{sec:dust_opacity} describes the main properties of the dust models used to derive the dust opacities.

\subsection{Disk structure}
\label{sec:disk_structure}

We start by defining an ``unperturbed'' disk structure on the top of which we will add the contribution of optically thick disk regions. 
For a disk with midplane temperature $T_{\rm{mid}}(r)$ and dust mass surface density $\Sigma(r)$ between an inner and outer radius $R_{\rm{in}}$ and $R_{\rm{out}}$, respectively, the sub-mm/mm SED can be modelled as the sum of the contribution by infinitesimal annuli with radius $r$ (in cylindrical coordinate system):  

\begin{equation}
F_{\nu}^{\rm{unp}} = \frac{\cos i}{d^{2}}\int_{R_{\rm{in}}}^{R_{\rm{out}}}B_{\nu}(T_{\rm{mid}}(r)) \left( 1-e^{-\tau_{\nu}(r)}\right)2\pi r dr,
\label{eq_unp_disk_flux}
\end{equation}
where $i$ is the disk inclination, i.e. the angle between the disk axis and the perpendicular to the plane of the sky\footnote{This parameter does not affect the sub-mm/mm significantly as long as the emission at these wavelengths is dominated by the optically thin disk regions. However this it not true anymore when emission from optically thick regions becomes important. In this paper we have adopted a value of 30\degree for the disk inclination of all the modelled disks.}, $d$ is the distance of the disk from the observer, and $B_{\nu}$ is the Planck function. In Eq.~\ref{eq_unp_disk_flux} the optical depth $\tau_{\nu}(r)$ is defined as

\begin{equation}
\tau_{\nu}(r)=\frac{\kappa_{\nu}\Sigma_{\rm{dust}}(r)}{\cos i},
\label{eq_tau}
\end{equation}
where $\kappa_{\nu}$ is the dust opacity coefficient\footnote{Note that here we have implicitly assumed that the dust opacity does not change within the disk.} (Section~\ref{sec:dust_opacity}), and $\Sigma_{\rm{dust}}(r)$ is the dust surface density.

In this work we adopted a modified
version of the two-layer models of passively irradiated
flared disks developed by Dullemond et al.~(\cite{Dul01}, which follows
the schematization by Chiang \& Goldreich~\cite{Chi97}) to derive the disk structure. In order for these models to calculate the temperature of the disk midplane, a surface
density profile has to be given as input. 

For the radial profile of the dust surface density we consider here the self-similar solution for a viscous disk (Lynden-Bell \& Pringle~\cite{Lyn74}):

\begin{equation}
\Sigma_{\rm{dust}}(r)=\Sigma_{0}\left(\frac{r}{r_{c}}\right)^{-\gamma}\exp\left[-\left(\frac{r}{r_{c}}\right)^{2-\gamma}\right],
\label{eq_sigma}
\end{equation}
with values of the $\Sigma_{0}$, $\gamma$, $r_{c}$ parameters in the ranges observationally constrained by sub-arcsec angular resolution imaging of young protoplanetary disks in the sub-mm (Isella et al.~\cite{Ise09}, Andrews et al.~\cite{And09},~\cite{And10}; see Section~\ref{sec:results}).

After setting the dust surface density, the properties
of the dust grains and of the central star (see below), the temperature profile $T_{\rm{mid}}(r)$ is derived by balancing the heating due to the reprocessed stellar
radiation by the disk surface layers with cooling due to the
dust thermal emission.
Here we consider a PMS star with a mass of $0.5~M_{\odot}$, bolometric luminosity of $0.9~L_{\odot}$ and effective temperature of $4000~$K, which are typical values for the sample of low-mass PMS stars studied in Ricci et al.~(\cite{Ric10a},~\cite{Ric10b},~\cite{Ric11a},~\cite{Ric11b}) in the Taurus, Ophiuchus and Orion Nebula Cluster forming regions.

On the top of the ``unperturbed'' disk structure just defined, we 
add the contribution of optically thick regions in the disk, i.e. regions where the optical depth $\tau_{\nu}>>1$ even at sub-mm and mm wavelengths. By noticing that for a completely optically thick disk the Eq.~\ref{eq_unp_disk_flux} for the flux density reduces to 

\begin{equation}
F_{\nu}^{\rm{thick}} = \frac{\cos i}{d^{2}}\int_{R_{\rm{in}}}^{R_{\rm{out}}}B_{\nu}(T_{\rm{mid}}(r)) 2\pi r dr,
\label{eq_disk_flux_opt_thick}
\end{equation}
we modify Eq.~\ref{eq_unp_disk_flux} by adding at each radius $r$ a fraction $f$ of the area of the disk that is optically thick, so that the total flux density becomes:

\begin{eqnarray}
% e^x &\approx& 1+x+x^2/2! + \\
%   && {}+x^3/3! + x^4/4! + \\
%   && + x^5/5!
F_{\nu}^{f} &=& \frac{\cos i}{d^{2}}\int_{R_{\rm{in}}}^{R_{\rm{out}}}B_{\nu}(T_{\rm{mid}}(r))\left[ f+(1-f)\left( 1-e^{-\tau_{\nu}(r)}\right)\right]\times \nonumber \\
&& {}\times 2\pi r dr
\label{eq_disk_flux}
\end{eqnarray}

The ``filling factor'' $f$ can thus be read as the fraction of area in the disk which is optically thick at all the wavelengths considered in the analysis. In general $f=f(r)$, i.e. the filling factor is a function of radius. If $f=0$ at all radii, Eq.~\ref{eq_disk_flux} reduces to Eq.~\ref{eq_unp_disk_flux}, i.e. no optically thick regions are added to the flux density of an unperturbed disk. If $f=1$ at all radii, Eq.~\ref{eq_disk_flux} reduces to Eq.~\ref{eq_disk_flux_opt_thick} for the flux density of a completely optically thick disk.

In Section~\ref{sec:f_const_F_alpha} we will treat firstly the simplest case of $f(r)$ constant with radius, then we will consider for $f(r)$ a family of step functions which are non-zero only within specified regions in the disk in Sect. \ref{sec:f_var}.   

Note that in Eq.~\ref{eq_disk_flux_opt_thick} and~\ref{eq_disk_flux} we assumed for simplicity that the temperature of the optically thick regions at a stellocentric radius $r$ is equal to the one of the unperturbed disk calculated by the two-layer disk model at the same radius. Although not strictly correct, this is a very good approximation. The temperature of the disk is roughly proportional to the ratio between the optical depth in the optical (where the disk absorbs most of the radiation energy) and the optical depth in the infrared (where the disk emits most of its radiation; see Dullemond et al.~\cite{Dul01}). Therefore, at first order, the dust temperature does not depend on the absolute value of the optical depth. 

Since for nearly all the disks imaged at high angular resolution in the sub-mm $\gamma$ is larger than 0 (Andrews et al.~\cite{And09},~\cite{And10}, Isella et al.~\cite{Ise09}), Eq.~\ref{eq_sigma} implies that the dust surface density typically decreases with distance from the central star. This means that the innermost regions of unperturbed disks can be dense enough to be optically thick at long mm wavelengths, as it was recognized since the very first analyses of sub-mm observations of YSOs (e.g. Beckwith et al.~\cite{Bec90}). Testi et al.~(\cite{Tes01}) quantified the impact of optically thick emission from inner disk regions by showing how the predicted mm-fluxes change when considering different parameters for the disk structure. To visualize their main results, they used the $F_{\rm{mm}}$ vs $\alpha_{\rm{mm}}$ diagram, where $F_{\rm{mm}}$ is the flux-density at a given mm-wavelength and $\alpha_{\rm{mm}}$ is the mm-spectral index. In this paper we will extensively use this diagram. The main difference with this and other previous works is that our attention will mostly focus on the impact of possible optically thick regions throughout the disk, rather than only on the optically thick inner regions in the unperturbed disk structure.
A parametric study of the effect of these local optically thick regions on the mm-SED, and of their properties is presented in Section~\ref{sec:results}. 
The physical origin of these regions is discussed in Section~\ref{sec:plausibility}.

\subsection{Dust opacity}
\label{sec:dust_opacity}

The only physical quantity present in Eq.~\ref{eq_unp_disk_flux}, \ref{eq_tau}, \ref{eq_disk_flux} which is left to be described is the frequency-dependent dust opacity coefficient $\kappa_{\nu}$. 
This term, which at long wavelengths represents the level of emissivity of the disk per unit dust mass, can be specified only when a model for the dust grain chemical composition, porosity, shape and size is considered (see e.g. Natta et al.~\cite{Nat04}).  

In this paper we adopt the same dust grain model as in Birnstiel et al.~(\cite{Bir10a}), Ricci et al.~(\cite{Ric10a}, \cite{Ric10a}), i.e. porous composite spherical grains made of astronomical silicates, carbonaceous materials and water ice (optical constants from Weingartner \& Draine~\cite{Wei01}, Zubko et al.~\cite{Zub96} and Warren \cite{War84}, respectively). The refractive indices of the different species have been combined by using the Bruggeman mixing theory. We use the fractional abundances estimated by Semenov et al.~(\cite{Sem03}).
%, and we consider three different grain porosities for our disk models (10\%, 30\% and 50\% in volume fraction).
After setting the chemical composition, porosity and shape of the grain, a grain size distribution $n(a)$ has to be specified in order to determine the dust opacity coefficient $\kappa_{\nu}$ of the dust model. We adopt a power-law grain size distribution with power-law exponent $q$

\begin{equation}
n(a) \propto a^{-q}
\label{eq_n_a}
\end{equation}
truncated between the minimum and maximum grain sizes $a_{\rm{min}}$ and $a_{\rm{max}}$, respectively.
Since in this work we want to investigate the impact of local optically thick regions in the outer disk on the slope of the mm-SED \textit{without invoking the presence of large mm-sized grains}, we consider here a size distribution which does not contain mm-grains, i.e. with $a_{\rm{max}} \approx 0.1$~mm. 
%has been proposed by Mathis et al.~(\cite{Mat77}) to fit extinction measurements of the interstellar medium. For this ISM-like dust $a_{\rm{max}} \approx 0.1~\mu$m, $a_{\rm{min}} << a_{\rm{max}}$, and $q \approx 3.5$. 
Note that, for any given grain chemical composition, as long as $a_{\rm{max}} \simless 0.1$~mm the dust opacity coefficient $\kappa_{\nu}$ at mm wavelengths stays unchanged and does not even depend on $q$~(see Fig.~3 in Ricci et al.~\cite{Ric10a}). Therefore, the dust model adopted here refers to a more general dust population in which only grains smaller than $\sim 0.1$~mm are present.

\section{Results}
\label{sec:results}

In this section we present the results of our analysis. In particular, in Sect.~\ref{sec:f_const_F_alpha} and \ref{sec:f_var} we compare the mm-flux densities as derived by our disk models for different values of parameters related to the disk structure, with those measured by Ricci et al.~(\cite{Ric10a}, \cite{Ric10b}, \cite{Ric11a}, \cite{Ric11b}) for 49 disks in the Taurus, Ophiuchus and Orion star forming regions. 

\subsection{Case of a constant $f$: the $F_{\rm{1mm}}$-$\alpha_{\rm{1-3mm}}$ diagram}
\label{sec:f_const_F_alpha}

\begin{figure*}[htbp!]
%\vspace{-1cm} 
%\hspace{-1.2cm}
 \centering
 \includegraphics[height=135mm]{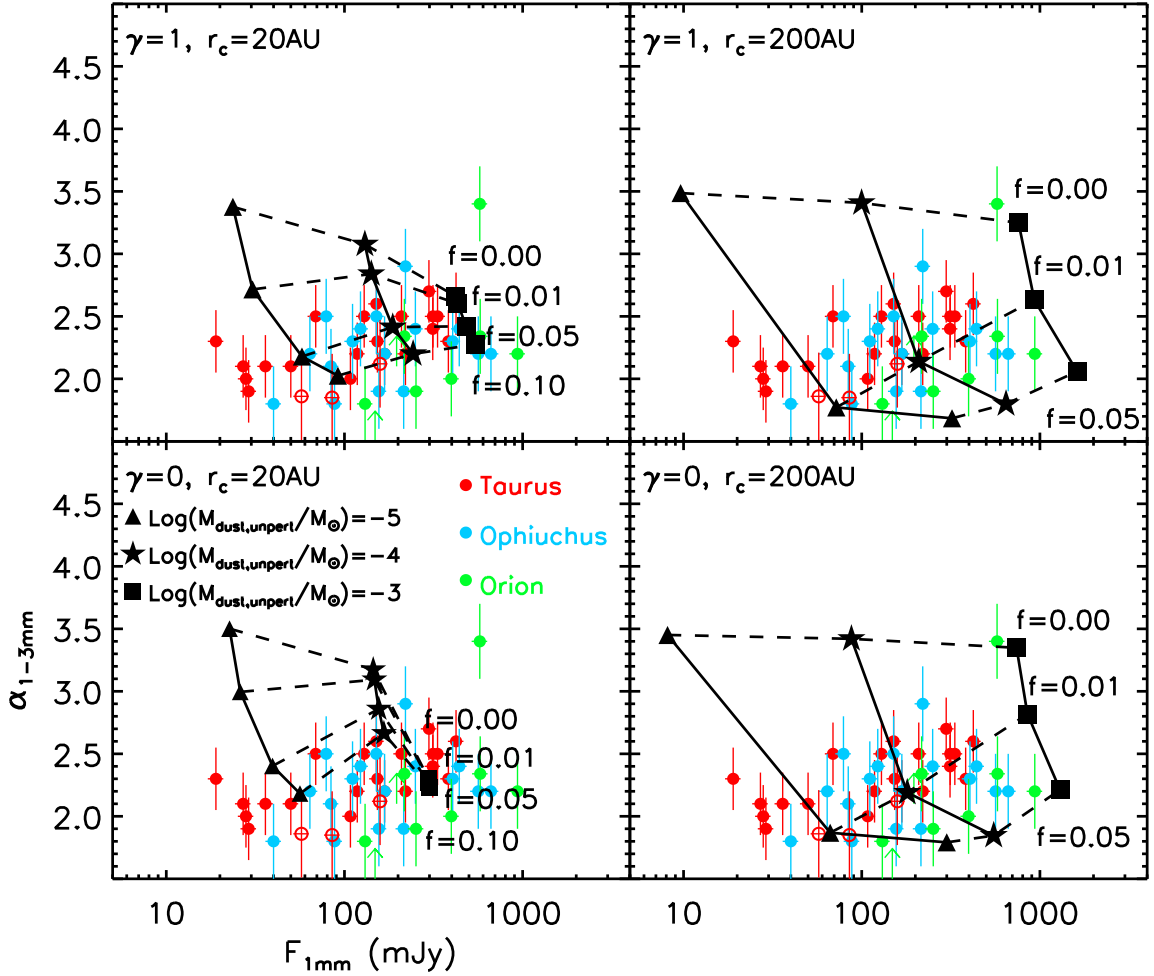}
% \vspace{-4.5cm}
 \caption{Flux density at 1mm vs spectral index between 1 and 3mm. In each panel red, blue and green points represent observational data of disks in Class II YSOs in Taurus (Ricci et al.~\cite{Ric10a}, and this work), Ophiuchus (Ricci et al.~\cite{Ric10b}), and Orion Nebula Cluster~(Ricci et al.~\cite{Ric11a},~\cite{Ric11b}), respectively. The 1~mm-fluxes in the Orion Nebula Cluster have been multiplied by a factor of (420 pc/140 pc)$^2$ to compensate for the different distances estimated for the Orion ($\sim$ 420~pc, Menten et al.~\cite{Men07}) and Taurus/Ophiuchus regions ($\sim$ 140~pc, Bertout et al.~\cite{Ber99}, Wilking et al.~\cite{Wil08}). Red empty circles are the three Class II YSOs observed with CARMA and discussed in the Appendix A. Black symbols show the model predictions for different unperturbed dust mass: triangles for $10^{-5}~M_{\odot}$, stars for $10^{-4}~M_{\odot}$, and squares for $10^{-3}~M_{\odot}$. Solid lines connect disk models with the same unperturbed dust mass, whereas dashed lines connect models with the same constant filling factor $f$ for the optically thick regions, with values indicated in the bottom right side of each panel. In each panel a given couple of the ($\gamma$, $r_{c}$)-parameters has been assumed for the unperturbed disk structure, with values indicated in the top left corner.}
\label{fig:flux_alpha_fconst_vac30} 
 \end{figure*}

We start by considering a $f(r)$-function which is constant for stellocentric radii lower than 300~AU\footnote{For all the disks considered in this work, more than 80\% of the unperturbed disk mass is contained within 300~AU.} and equal to 0 outside 300~AU.
Each panel in Figure~\ref{fig:flux_alpha_fconst_vac30} shows the prediction of our disk models for different values of the filling factor $f$ and of the dust mass for a given couple of the ($\gamma$,$r_c$) parameters for the disk unperturbed structure\footnote{The value of the normalization factor in equation~\ref{eq_sigma}, i.e. $\Sigma_0$, is set after fixing the (unperturbed) dust mass in the disk.}. The values of the parameters considered in this work for the unperturbed disk, i.e. $\gamma=0, 1$ and $r_c = 20, 200$~AU (see equation~\ref{eq_sigma}), lay at the limits of the ranges for these parameters as recently constrained by a high-angular resolution survey of about 30 protoplanetary disks in the sub-mm (Isella et al.~\cite{Ise09}, Andrews et al.~\cite{And09},~\cite{And10}). 
%The impact of the variation of $\gamma$ and $r_{c}$ is investigated later on. 
%The dust masses of the unperturbed disks considered in this paper are shown in the same figure.  

In each panel the points with $f=0$ represent the emission of the unperturbed disk, i.e. without any addition of optically thick regions. The dependence of the flux density at 1~mm on the dust mass is due to the fact that the bulk of the material reside in the outer disk regions where the surface density is relatively low and the emission optically thin. In terms of the spectral index $\alpha_{\rm{1-3mm}}$ the points with $f=0$ lay at relatively large values of about $3-3.5$. This is due to the adopted dust model made of small grains (Section~\ref{sec:dust_opacity}), for which the value of the spectral slope $\beta$ of the dust opacity coefficient is about 1.5. The slight decrease of $\alpha_{\rm{1-3mm}}$ with increasing dust mass is given by the fact that for disks that are massive enough the innermost disk regions ($r < 10-20$~AU) are so dense that the emission from these regions becomes optically thick in the sub-mm, and this has the effect of making the SED shallower. 

Note however that the points with $f=0$ do not get into the area of the $F_{\rm{1mm}}$-$\alpha_{\rm{1-3mm}}$ diagram which contains the bulk of the observational data. In particular the models presented here with $f=0$ overpredict the observed spectral index $\alpha_{\rm{1-3mm}}$. 
%In the literature, the commonly adopted interpretation for this discrepancy between models and data is that the assumption of a ISM-like distribution of particle sizes in disks is wrong. By considering in the optically thin outer disk dust particles larger than about 1~mm, the spectral index $\alpha_{\rm{1-3mm}}$ gets lower because of a decrease in the spectral index $\beta$ of the dust opacity coefficient as compared with the value for the ISM-like dust.
%In this work, we investigate the feasibility of an alternative scenario in which the measured low spectral indices are due to optically thick regions in the outer disk rather than to the presence of large grains in the outer disk, as detailed in the last Section.     
%These optically thick regions might be caused by an increase of the optical depth as due to the local concentration of particles. Different physical processes that have the potential of concentrate particles have been proposed in the literature with the aim of investigating the formation of planetesimals (). 
%In this Section we want to quantify the effect of these possible optically thick regions 

The effect of increasing the filling factor $f$ of optically thick regions has always the same kind of signature in the ($F_{\rm{1mm}}$,$\alpha_{\rm{1-3mm}}$)-plane: the absolute flux in the (sub-)mm increases because more and more optical depth is added into the system, whereas the spectral index tends to decrease and approach the value of about 2, as expected for completely optically thick emission in the Rayleigh-Jeans regime\footnote{Note that in some cases the disk models predict values for the spectral index $\alpha_{\rm{1-3mm}}$ which are lower than 2. This is because the Rayleigh-Jeans regime is not completely reached at these wavelengths in the outer and cooler regions of the disks.}.
  
The main result shown in Figure~\ref{fig:flux_alpha_fconst_vac30} is that, for most of the unperturbed disk structures, relatively low values of the filling factor ($f \simless 0.05$) are required to explain the bulk of the data.
The most significant variation of the mm-fluxes with the properties of the unperturbed disk comes when the characteristic radius changes from 20 to 200~AU (see left and right panels, respectively). 
With $r_c = 20$~AU the surface density is more concentrated in the inner disk than in the $r_c = 200$~AU case. As a consequence of this the impact of the optically thick inner regions to the total emission is more evident (see e.g. the low $\alpha_{\rm{1-3mm}}$-value of the model with $f=0$ and a dust mass of $10^{-3}~M_{\odot}$ for the unperturbed disk structure). However, the fluxes of the disks with $f > 0$ do not dramatically depend on the structure of the unperturbed disk. 
This is due to the fact that in most of the cases already for $f \approx 0.05$ the total disk emission becomes dominated by the added optically thick regions.
 
What varies strongly with the unperturbed disk structure at a given $f$ is the mass which has to be included in these added regions in order to make them optically thick. This is discussed in the next section.     

%\begin{figure}[tbp!]
%%\vspace{-1cm} 
%%\hspace{-1.2cm}
% \centering
% \includegraphics[scale=0.5]{figures/flux_alpha_fconst_vac30_g1_r60.eps}
%% \vspace{-4.5cm}
% \caption{ }
%\label{fig:flux_alpha_fconst_vac30_g1_r60} 
% \end{figure}
%
%\begin{figure}[tbp!]
%%\vspace{-1cm} 
%%\hspace{-1.2cm}
%% \centering
% \includegraphics[scale=0.5]{figures/flux_alpha_fconst_vac30_g1_r20.eps}
%% \vspace{-4.5cm}
% \caption{Same as in Figure~\ref{fig:flux_alpha_fconst_vac30_g1_r60}, but with a characteristic radius $r_c$ of 20~AU.}
%\label{fig:flux_alpha_fconst_vac30_g1_r20} 
% \end{figure}
%
%\begin{figure}[tbp!]
%%\vspace{-1cm} 
%%\hspace{-1.2cm}
% \centering
% \includegraphics[scale=0.5]{figures/flux_alpha_fconst_vac30_g1_r200.eps}
%% \vspace{-4.5cm}
% \caption{Same as in Figure~\ref{fig:flux_alpha_fconst_vac30_g1_r60}, but with a characteristic radius $r_c$ of 200~AU.}
%\label{fig:flux_alpha_fconst_vac30_g1_r200} 
% \end{figure}

\subsection{Case of a constant $f$: required masses in optically thick regions}
\label{sec:f_const_req_overd}

So far we have considered the emission of our modelled disks without considering how much mass has to be added to the disk in order to make a fraction of the disk surface optically thick even at millimeter wavelengths.
A simple way to do this is to calculate the surface density which has to be present in the added regions 
to give them an optical depth of about 1 at the longest wavelength considered in this work, i.e. 3~mm, and then integrate over the disk area covered by the filling factor $f$. This represents actually a lower limit for the mass which has to be present in the added regions to make them optically thick: if more mass is put onto those regions this does not have a significant effect onto the SED.

Therefore, the surface density of dust in the optically thick regions $\Sigma_{\rm{dust}}^f$ is given by the condition

\begin{equation}
\tau_{\rm{3mm}} \approx \Sigma_{\rm{dust}}^f \kappa_{\rm{3mm}} \approx 1  \rightarrow \Sigma_{\rm{dust}}^f \approx 1/\kappa_{\rm{3mm}},
\label{eq:Sigma_f}
\end{equation}

\noindent which implies that $\Sigma_{\rm{dust}}^f$ does not depend on the stellocentric radius. As a consequence of this the total amount of dust in these regions

\begin{equation}
M_{\rm{dust}}^f = f \int_{R_{\rm{in}}}^{R_{\rm{out}}} \Sigma_{\rm{dust}}^f 2 \pi r dr \approx f \Sigma_{\rm{dust}}^f \pi R_{\rm{out}}^2, 
\label{eq:dust_mass_f}
\end{equation}

\noindent i.e. the total dust mass in the optically regions depends quadratically on the largest radius in which these regions are present in the disk, namely 300~AU in this simulation. For the dust considered in Section~\ref{sec:dust_opacity}, $\kappa_{\rm{3mm}} \approx 0.45$~cm$^2$/g and $M_{\rm{dust}}^f \approx f \times 0.07~M_{\odot}$. This means that even in the case of the model with the largest unperturbed mass in dust ($10^{-3}~M_{\odot}$) and with the lowest value of the filling factor $f$ considered here (0.01), the optically thick regions contain at least 70\% of the mass in the unperturbed disk. This ratio then increases linearly with $f$ and decreases with increasing the dust mass of the unperturbed disk. This argument clearly shows that in order to have optically thick regions of the kind discussed so far in the outer disk a very strong redistribution of dust particles has to occur in the disk (see Section~\ref{sec:discussion}).  

As shown in Eq.~\ref{eq:dust_mass_f}, the dust mass in the optically thick regions depends quadratically on the disk outer radius. An idea to reduce the amount of dust required in these optically thick regions is therefore to reduce the area in the disk where they are located. This is the topic of the next section.

\subsection{Case of $f(r)$ as a step function: the $F_{\rm{1mm}}$-$\alpha_{\rm{1-3mm}}$ diagram}
\label{sec:f_var}

We consider here the result of optically thick regions localized in smaller areas of the disk. In particular, we discuss four cases in which $f(r)$ is a step function with values greater than 0 between 10 and 20~AU, 30 and 40~AU, 50 and 60~AU, 80 and 90~AU, respectively. We consider annuli with a width of $10$~AU because the physical mechanisms proposed so far to concentrate particles locally in the disk typically act on these length scales or smaller (see Section~\ref{sec:plausibility}).   
The different central radii chosen for the annuli allow us to investigate how the location of the optically thick regions in the disk can affect its total sub-mm emission.

\begin{figure*}[htbp!]
%\vspace{-1cm} 
%\hspace{-1.2cm}
 \centering
 \includegraphics[scale=1.0]{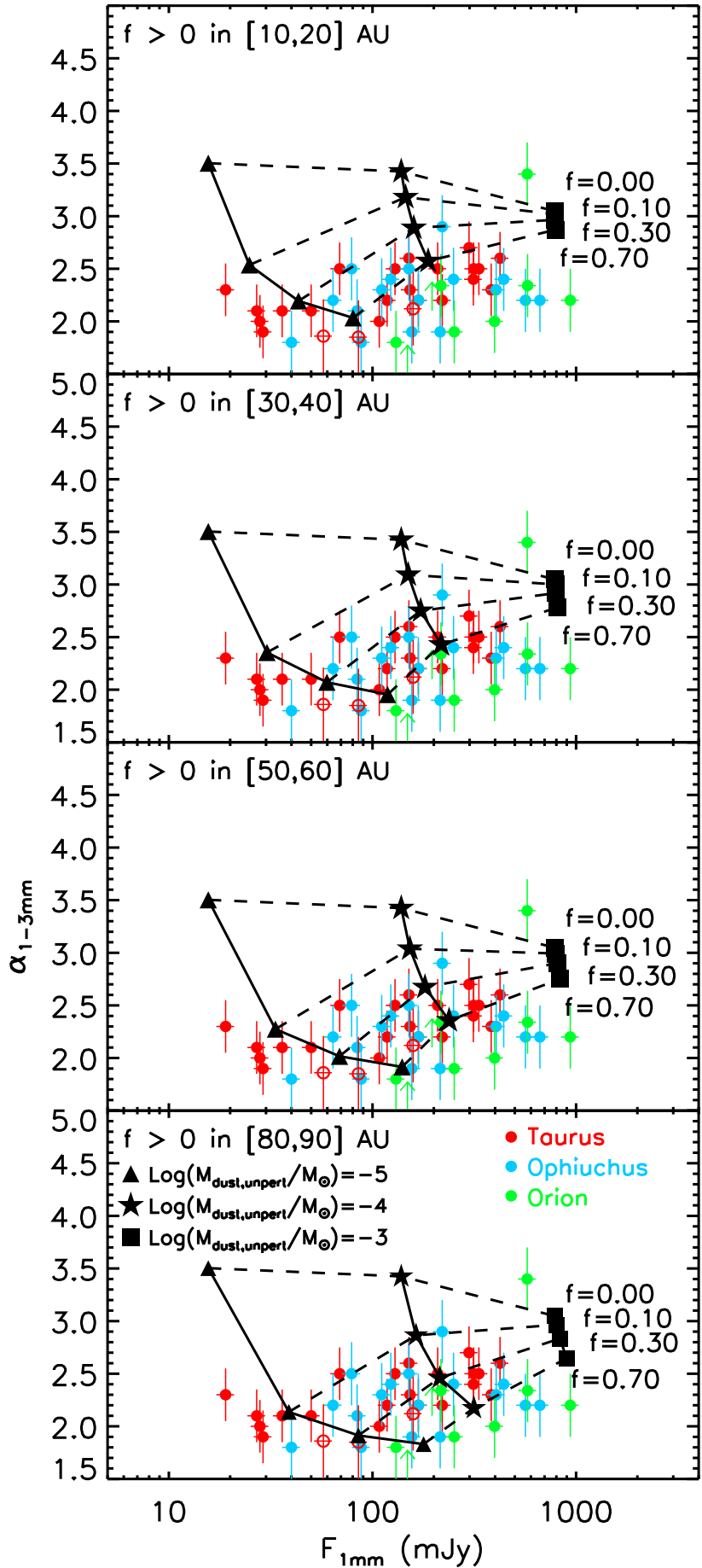}
% \vspace{-4.5cm}
 \caption{As in Fig.~\ref{fig:flux_alpha_fconst_vac30}, but with optically thick regions localized in annuli of 10~AU-width, with inner and outer radii indicated in the top left corner of each panel. The unperturbed structure of the disk is characterized by $\gamma = 0.5$ and $r_c = 60$~AU.}
\label{fig:flux_alpha_fvar_gamma05_rc60} 
 \end{figure*}

The four panels in Fig.~\ref{fig:flux_alpha_fvar_gamma05_rc60} show the model predictions on the $F_{\rm{1mm}}-\alpha_{\rm{1-3mm}}$ diagram for disks with such localized optically thick regions and an unperturbed disk structure with $\gamma = 0.5$ and $r_c = 60$~AU, which are approximately the median values for disks imaged at high-angular resolution.   

The optically thick regions can have a signficant impact onto the global mm-SED even if they are concentrated in rings with the relatively small width of 10~AU. The effect is stronger for the disk with lower masses because of the higher constrast in optical depth between the unperturbed disk and the optically thick regions. For the same reason, i.e. the contrast with the unperturbed disk structure, and for the fact that the area of an annulus of a given width scales linearly with the central radius of the annulus, at a given filling factor $f$ inside the ring, the effect of the optically thick regions is the largest in the ring which is the furthest from the star, i.e. between 80 and 90~AU.
Note that to reproduce the bulk of the data, larger fractions $f$ are needed than in the $f(r) =$ const case. This is because in the case of optically thick emission, apart for the temperature, it is the surface area of the emitting material that determines the amount of its emission. Therefore, if these regions are distributed over a smaller area of the disk, they need to occupy a larger fraction of that area, which is what we find with our analysis.      
This argument justifies our choice of dealing with very simple structures for the optically thick regions: even if optically thick regions in real disks would likely have more complex structures than modelled here, our analysis is meaningful in terms of the fractional area covered by     
such regions throughout the disk.

%
%\begin{figure}[tbp!]
%%\vspace{-1cm} 
%%\hspace{-1.2cm}
% \centering
% \includegraphics[scale=0.5]{figures/flux_alpha_f80_90_vac30_g1_r60.eps}
%% \vspace{-4.5cm}
% \caption{Same as in Figure~\ref{fig:flux_alpha_fconst_vac30_g1_r60}, but .}
%\label{fig:flux_alpha_f80_90_vac30_g1_r60} 
% \end{figure}
%
%\begin{figure}[tbp!]
%%\vspace{-1cm} 
%%\hspace{-1.2cm}
% \centering
% \includegraphics[scale=0.5]{figures/flux_alpha_f50_60_vac30_g1_r60.eps}
%% \vspace{-4.5cm}
% \caption{Same as in Figure~\ref{fig:flux_alpha_fconst_vac30_g1_r60}, but .}
%\label{fig:flux_alpha_f50_60_vac30_g1_r60} 
% \end{figure}
%  
%\begin{figure}[tbp!]
%%\vspace{-1cm} 
%%\hspace{-1.2cm}
% \centering
% \includegraphics[scale=0.5]{figures/flux_alpha_f10_20_vac30_g1_r60.eps}
%% \vspace{-4.5cm}
% \caption{Same as in Figure~\ref{fig:flux_alpha_fconst_vac30_g1_r60}, but .}
%\label{fig:flux_alpha_f10_20_vac30_g1_r60} 
% \end{figure}

\subsection{Case of $f(r)$ as a step function: required overdensities}
\label{sec:overdensities_f_var}

As done in Section~\ref{sec:f_const_req_overd} for the case of a constant $f$ within 300~AU, we analize here the dust mass which has to be present in the optically thick regions.
This can be calculated by using the first equality in Eq.~\ref{eq:dust_mass_f} with the inner and outer radii of the ring for $R_{\rm{in}}$ and $R_{\rm{out}}$, respectively.

For the case of the ring between 10 and 20~AU the total dust mass in the optically thick regions is $M_{\rm{dust}}^{f} \approx f \times 2.4 \cdot 10^{-4}~M_{\odot}$, and rises to $\approx f \times 5.5 \cdot 10^{-4}~M_{\odot}$, $\approx f \times 8.7 \cdot 10^{-4}~M_{\odot}$ and $\approx f \times 1.3 \cdot 10^{-3}~M_{\odot}$ when the ring is moved outward to 50-60~AU and 80-90~AU, respectively, because of the increased area of the ring itself.
To explain the left end of the $F_{\rm{1mm}}$-$\alpha_{\rm{1-3mm}}$ diagram (Fig.~\ref{fig:flux_alpha_fvar_gamma05_rc60}) filling factors $f \simgreat 0.1$ on the top of the lowest mass disk ($M_{\rm{dust,unpert}} = 10^{-5}~M_{\odot}$) are needed. However, the mass in the optically thick regions is larger than the one in the unperturbed structure for all the rings considered here. This would require an extremely strong concentration of particles in those regions which are not seen in the results of the numerical simulations run so far (see discussion in Section~\ref{sec:plausibility}).    
For the disk with $M_{\rm{dust,unpert}} = 10^{-4}~M_{\odot}$ a significant decrease of the spectral index $\alpha_{\rm{1-3mm}}$, i.e. down to about 2.5 and lower is obtained only for $f \simgreat 0.3$ (Fig.~\ref{fig:flux_alpha_fvar_gamma05_rc60}). These filling factors require dust masses in the optically thick regions as large as at least 70\% of the unperturbed disk mass.     
Finally, the added optically thick regions in the most massive disk considered here, with $M_{\rm{dust,unpert}} = 10^{-3}~M_{\odot}$, contain relatively low dust mass as compared with the unperturbed disk mass. For example, in the case of $f=0.5$ the ratio among the former and latter masses is about 12\%, 28\%, 44\%, 65\% for the optically thick regions inside rings with radii of 10-20~AU, 30-40~AU, 50-60~AU and 80-90~AU, respectively.

Tables~\ref{tab:overd_10_20AU}-\ref{tab:overd_80_90AU} report, for models with different unperturbed disk structures in terms of total dust mass, $\gamma$ and $r_{\rm{c}}$, the required overdensities in the optically thick regions. These are defined as the ratio between the dust surface density $\Sigma_{\rm{dust}}^{\rm{f}}$ in the optically thick regions and the surface density in the unperturbed disk structure evaluated at the center of the ring. Since $\Sigma_{\rm{dust}}^{\rm{f}}$ is always the same (i.e. 1/$k_{\rm{3mm}} \approx 2.2$~cm$^2$/g), the required overdensities depend only on the value that the dust surface density of the unperturbed disk assumes in the ring, and therefore on the ($M_{\rm{dust,unpert}}, \gamma, r_{c}$) parameters which define such surface density. Note that the values of these overdensities do not depend on the adopted value for the dust opacity $\kappa_{\rm{3mm}}$ (or equivalently on the adopted dust model). This is because both the surface density in the optically thick regions, and the surface density in the optically thin unperturbed disk structure at a given flux depend on the dust opacity as $\kappa_{\rm{3mm}}^{-1}$. The lowest values for the overdensities are found for the most massive disk when the overdensity region is located closest to the star. The low overdensity values of order of unity indicate that this disk is massive enough to have the unperturbed structure at radii of $\sim$10-20~AU marginally optically thick by itself, i.e. without the addition of any artificial optically thick region.

\vspace*{3mm}

\begin{table*}
\centering \caption{required dust overdensities in the added optically thick regions inside a ring between 10 and 20~AU from the central star.}

\begin{tabular}{c|c|c|c|c|c|c|}
\cline{2-7}
& \multicolumn{2}{|c|}{$\gamma = 1$} & \multicolumn{2}{|c|}{$\gamma = 0.5$} & \multicolumn{2}{|c|}{$\gamma = 0$} \\ \cline{2-7}
& $r_{\rm{c}}=20$~AU & $r_{\rm{c}}=200$~AU & $r_{\rm{c}}=20$~AU & $r_{\rm{c}}=200$~AU  & $r_{\rm{c}}=20$~AU & $r_{\rm{c}}=200$~AU \\ \cline{1-7}
%\multicolumn{1}{|c|}{\multirow{3}{*}{$M_{\rm{disk,unperturb}} = 10^{-5}~M_{\odot}$}} &
\multicolumn{1}{|c|}{$M_{\rm{dust,unperturb}} = 10^{-5}~M_{\odot}$} & 100 & 520 & 70 & 1200  & 55 & 3200 \\ \cline{1-7}
\multicolumn{1}{|c|}{$M_{\rm{dust,unperturb}} = 10^{-4}~M_{\odot}$} & 10 & 52   & 7 & 120 & 5.5 & 320 \\ \cline{1-7}
\multicolumn{1}{|c|}{$M_{\rm{dust,unperturb}} = 10^{-3}~M_{\odot}$} & 1 & 5.2   & 0.7 & 12 & 0.55 & 32  \\ \cline{1-7}

\end{tabular}
\label{tab:overd_10_20AU}
\end{table*}

\vspace*{3mm}

\begin{table*}
\centering \caption{required dust overdensities in the added optically thick regions inside a ring between 30 and 40~AU from the central star.}

\begin{tabular}{c|c|c|c|c|c|c|}
\cline{2-7}
& \multicolumn{2}{|c|}{$\gamma = 1$} & \multicolumn{2}{|c|}{$\gamma = 0.5$} & \multicolumn{2}{|c|}{$\gamma = 0$} \\ \cline{2-7}
& $r_{\rm{c}}=20$~AU & $r_{\rm{c}}=200$~AU & $r_{\rm{c}}=20$~AU & $r_{\rm{c}}=200$~AU  & $r_{\rm{c}}=20$~AU & $r_{\rm{c}}=200$~AU \\ \cline{1-7}
%\multicolumn{1}{|c|}{\multirow{3}{*}{$M_{\rm{disk,unperturb}} = 10^{-5}~M_{\odot}$}} &
\multicolumn{1}{|c|}{$M_{\rm{dust,unperturb}} = 10^{-5}~M_{\odot}$} & 630 & 1300 & 530 & 2000  & 680 & 3300 \\ \cline{1-7}
\multicolumn{1}{|c|}{$M_{\rm{dust,unperturb}} = 10^{-4}~M_{\odot}$} & 63 & 130   & 53 & 200 & 68 & 330 \\ \cline{1-7}
\multicolumn{1}{|c|}{$M_{\rm{dust,unperturb}} = 10^{-3}~M_{\odot}$} & 6.3 & 13   & 5.3 & 20 & 6.8 & 33  \\ \cline{1-7}

\end{tabular}
\label{tab:overd_30_40AU}
\end{table*}

\begin{table*}
\centering \caption{required dust overdensities in the added optically thick regions inside a ring between 50 and 60~AU from the central star.}

\begin{tabular}{c|c|c|c|c|c|c|}
\cline{2-7}
& \multicolumn{2}{|c|}{$\gamma = 1$} & \multicolumn{2}{|c|}{$\gamma = 0.5$} & \multicolumn{2}{|c|}{$\gamma = 0$} \\ \cline{2-7}
& $r_{\rm{c}}=20$~AU & $r_{\rm{c}}=200$~AU & $r_{\rm{c}}=20$~AU & $r_{\rm{c}}=200$~AU  & $r_{\rm{c}}=20$~AU & $r_{\rm{c}}=200$~AU\\ \cline{1-7}
%\multicolumn{1}{|c|}{\multirow{3}{*}{$M_{\rm{disk,unperturb}} = 10^{-5}~M_{\odot}$}} &
\multicolumn{1}{|c|}{$M_{\rm{dust,unperturb}} = 10^{-5}~M_{\odot}$} & 2700 & 2300 & 5700 & 2700  & 62000 & 3400 \\ \cline{1-7}
\multicolumn{1}{|c|}{$M_{\rm{dust,unperturb}} = 10^{-4}~M_{\odot}$} & 270 & 230 & 570 & 270  & 6200 & 340 \\ \cline{1-7}
\multicolumn{1}{|c|}{$M_{\rm{dust,unperturb}} = 10^{-3}~M_{\odot}$} & 27 & 23   & 57  & 27  & 620 & 34 \\ \cline{1-7}

\end{tabular}
\label{tab:overd_50_60AU}
\end{table*}

\begin{table*}
\centering \caption{required dust overdensities in the added optically thick regions inside a ring between 80 and 90~AU from the central star.}

\begin{tabular}{c|c|c|c|c|c|c|}
\cline{2-7}
& \multicolumn{2}{|c|}{$\gamma = 1$} & \multicolumn{2}{|c|}{$\gamma = 0.5$}  & \multicolumn{2}{|c|}{$\gamma = 0$} \\ \cline{2-7}
& $r_{\rm{c}}=20$~AU & $r_{\rm{c}}=200$~AU & $r_{\rm{c}}=20$~AU & $r_{\rm{c}}=200$~AU  & $r_{\rm{c}}=20$~AU & $r_{\rm{c}}=200$~AU \\ \cline{1-7}
%\multicolumn{1}{|c|}{\multirow{3}{*}{$M_{\rm{disk,unperturb}} = 10^{-5}~M_{\odot}$}} &
\multicolumn{1}{|c|}{$M_{\rm{dust,unperturb}} = 10^{-5}~M_{\odot}$} & 19000 & 4100 & 370000 & 3800  & 2.4$\cdot 10^9$ & 3700 \\ \cline{1-7}
\multicolumn{1}{|c|}{$M_{\rm{dust,unperturb}} = 10^{-4}~M_{\odot}$} & 1900 & 410   & 37000 & 380  & 2.4$\cdot 10^8$ & 370 \\ \cline{1-7}
\multicolumn{1}{|c|}{$M_{\rm{dust,unperturb}} = 10^{-3}~M_{\odot}$} & 190 & 41     & 3700 & 38   & 2.4$\cdot 10^7$ & 37  \\ \cline{1-7}

\end{tabular}
\label{tab:overd_80_90AU}
\end{table*}

\section{Discussion}
\label{sec:discussion}

\subsection{Impact of the optically thick regions on high-angular resolution observations of disks in the sub-mm}
\label{sec:imaging}

So far we have discussed the possible effect of local optically thick regions in the \textit{integrated} fluxes of young disks in the millimeter. Since the required overdensities needed to make these regions optically thick are typically very large (especially for the low-mass disks, see Tables~\ref{tab:overd_10_20AU}-\ref{tab:overd_80_90AU}) one could expect these regions to be easily detectable with high-angular resolution imaging through (sub-)millimeter interferometry. However, the contrast between the surface brightnesses inside and outside the optically thick regions is not the only factor determining the possible observability of these structures. In fact, if the bright optically thick regions were uniformly distributed throughout the disk and with characteristic length scales much smaller than the angular resolution of the observations, they would be smeared out by the convolution with the resolution element. 

Since all the high-angular resolution observations of disks conducted so far in the (sub-)mm have revealed a disk structure which is essentially homogeneous, the results of existent observations provide an upper limit to the characteristic length scales of the invoked optically thick regions. The highest angular resolutions achieved so far in the sub-mm are about 0.2-0.4 arcsec, corresponding to physical scales of $\sim 30-60$~AU at the distances of nearby star forming regions (Isella et al.~\cite{Ise10}, Andrews et al.~\cite{And11}, Guilloteau et al.~\cite{Gui11}). Note that this is larger only by a factor of a few than the width of 20~AU considered in Section~\ref{sec:f_var}. This means that if optically thick regions were concentrated in those 20~AU-wide annuli, a radial bump in the surface brightness map could have been marginally detected, although not spatially resolved, by these observations.   
    
In order to probe non-homogeneous structures at smaller scales higher angular resolution is needed. The Atacama Large Millimeter/Submillimeter Array (ALMA) will allow to do that down to scales of a few AU only. For example, ALMA can detect spiral density waves in nearby massive disks~(Cossins et al.~\cite{Cos10}).
As detailed in the next subsection in these regions the concentration of particles might be efficient enough to make the dust emission from spiral arms optically thick.

If non-homogeneous regions will be detected by future observations with very high angular resolution, a prediction of the models presented in this work is that the millimeter spectral index measured in the bright regions should be equal to the spectral index of the Plank function in that region, which would be 2 in the case of Rayleigh-Jeans emission. To determine more precise predictions from these models would require further modelling of the disk structure with a proper treatment of the radiative transport in the optically thick regions. This goes well beyond the scope of this work. The future results coming from observations with ALMA and the EVLA arrays have the potential to guide future developments of disk models along these lines.

Recently, Guilloteau et al.~(\cite{Gui11}) presented an investigation of the radial variation of the dust properties through a dual-frequency (1.3 and 3~mm) survey of disks in the Taurus-Auriga region with the Plateau de Bure Interferometer. In most disks that are resolved at the two wavelengths, they find a radial dependence of the millimeter spectral index which they interpret as due to a variation of the spectral index of the dust opacity coefficient $\beta$ throughout the disk. In particular the $\beta$-index appears to typically increase from low values ($\sim$ 0-0.5) within $\sim 50-100$~AU from the central star to about 1.5-2 at the disk outermost regions. In the scenario presented in this work, the same data could be potentially explained in terms of a different efficiency of the physical mechanism(s) concentrating particles in different regions of the disk: in the inner disk, filling factors of optically thick regions $f \sim$ a few $\times~0.1$ could explain the lower spectral indices (see Fig.~\ref{fig:flux_alpha_fvar_gamma05_rc60}), thus mimicking the effect of a lower $\beta$;  
in the less dense outermost regions the same mechanisms might be not efficient enough to concentrate particles and that could explain the large spectral indices which are typical of optically thin emission from small dust grains (sizes $<$ 0.1~mm).

From the existent data only it is not possible to rule out the scenario described in this paper. 
We also note here that, if present, high levels of overdensities like those invoked by this analysis can probably occur only in the disk midplane. Because of the lower optical depths at longer wavelengths, only observations in the millimeter (or longer wavelengths) can probe directly the existence of these overdense regions. Observations in the near and mid-infrared probe only the surface layers of the disk, and they would be therefore insensitive to concentrations of particles in the midplane. Also observations in the far-IR ($\sim 100$~$\mu$m) are still significantly affected by optically thick emission from the unperturbed disk structure, at least for typical disks around T Tauri stars with disk masses larger than a few Jupiter masses (see e.g. the analysis by Andrews \& Williams~\cite{And05}). This would make the analysis presented in this paper much more model dependent at IR wavelengths than at longer mm-waves because of the impact of disk structure (e.g. flaring) and the particular choice for the treatment of the radiative transfer in the disk. Note that this might be not the case, especially in the far-IR, for disks with lower masses e.g. around young brown dwarfs. At the same time, depending on the physical mechanism driving this local concentration, the vertical distribution of particles in the disk, and therefore also in the surface layers, might be somewhat affected. In order to accurately determine this effect one would need global dynamical models of dust particles in a gaseous disk coupled with radiative transfer calculations to predict the disk emission in the IR.

In the next subsection we discuss the possible mechanisms proposed in the literature to concentrate particles in the disk. In particular we aim to understand how plausible are the overdensities required to explain optically thick regions considered in the last Sections. To do this we base the discussion on our current theoretical knowledge of the potential processes driving the concentration of solids in young disks.  

\subsection{Physical plausibility of the required overdensities}
\label{sec:plausibility}

To understand whether the overdense optically thick regions discussed in this paper are physically plausible in real protoplanetary disks, one has to compare the required filling factors and overdensities of regions with small $< 0.1$~mm-sized grains discussed in the last section with the outcome of the numerical simulations which investigate different mechanisms leading to the concentration of solid particles. These mechanisms are often invoked in the literature for their potential of halting locally the otherwise fast radial drift of solids (Brauer et al.~\cite{Bra07}) and of forming planetesimals in the disk (see a recent review by Chiang \& Youdin~\cite{Chi10}). 

A promising mechanism to concentrate solid particles is through the development of streaming instabilities, in which the gas and solid components are mutually coupled by drag forces in a turbulent disk. Johansen \& Youdin~(\cite{Joh07}) showed that overdensities of particles even larger than 1000 can be formed.
%\footnote{Note that in this paper the densities considered by the authors are in volume. Going from volume to surface densities is not straightforward, expecially considering the complexity of the derived structures and the fact that these are box-simulations, which therefore do not cover the full disk vertical extension. In this paper we will assume that volume and surface densities have the same order of magnitude.}. If converted to our definition of overdensities adopted here, which involve a peak-to-peak ratio rather than peak-to-average one as in their paper, values slightly larger than 1000 can be attained. 
Looking at Tables~\ref{tab:overd_10_20AU}-\ref{tab:overd_80_90AU} this indicates that these overdense regions would probably be optically thick also in the millimeter for most of the disk models considered here. However, in terms of the impact to the global disk SED, the filling factor $f$ of these regions is probably not large enough to be really significant: the fraction of particles in overdense regions at level of the order of 100 or above is only about 1\% or less (see Fig.~11 in Johansen \& Youdin~\cite{Joh07}). This means that the filling factor $f$ of disk area occupied by such regions is much lower than that, as the mean distance between particles decreases with increasing density.
Furthermore, all these highly-overdense regions are obtained for solid particles which are only marginally coupled to the gas, with values for the Stokes parameter\footnote{The Stokes parameter is an adimensional parametrization for the particle size in a gaseous disk. In the Epstein regime, which is relevant for the conditions in the disks treated in this work, St$=\pi a \rho_{\rm{s}}/(2\Sigma_{\rm{gas}})$, where $a$ and $\rho_{\rm{s}}$ are the particle size and density, respectively.},  St$\sim$1. Grains with sizes $< 0.1$~mm, as those considered in this analysis, are instead very well coupled to the gas (St$<<$1). In the case of streaming instabilities Johansen \& Youdin~(\cite{Joh07}) showed that already for particles with St$\sim$0.1, the level of overdensity is decreased by a factor of 10. For these reasons, we argue that, although streaming instabilities may potentially form overdense regions which are optically thick at millimeter wavelengths, they would not reach the required filling factor of the disk surface to explain the observations.             

Another possible mechanism which has been proposed to trap solids involves the presence of large anticyclonic vortices in the disk. These structures can be the result of baroclinic instability (Klahr \& Bodenheimer~\cite{Kla03}), the Rossby wave instability (Lovelace et al.~\cite{Lov99}, Regaly et al.~\cite{Reg11}), or magneto-rotational instability (Fromang et al.~\cite{Fro05}). Recent 2D numerical simulations of circumstellar disks have shown that, although dust overdensities larger than 2-3 order of magnitudes can be obtained in vortices covering a significant fraction of the disk surface area, the solids that are trapped are larger than about 10~cm, whereas much smaller sub-mm dust grains are not significantly affected by such structures (Lyra et al.~\cite{Lyr09}). 

Long-lived axisymmetric pressure bumps have been obtained in simulations of magneto-rotational turbulent disks (Johansen et al.~\cite{Joh09}). These pressure bumps have the potential of trapping solids which are marginally coupled to the gas, like mm-cm sized pebbles, but the obtained overdensities in gas are not large enough ($\simless 10-20\%$) to redistribute more gas-coupled small grains at the levels required for the optically thick regime (Tables \ref{tab:overd_10_20AU}
-\ref{tab:overd_80_90AU}).     
  
Finally, some concentration of dust particles can occur in disks undergoing gravitational instabilites.  
In these disks, the non-linear evolution of the instabilities lead to the formation of spiral waves with local overdensities in the gas component as high as about 100 (Rice et al.~\cite{Ric04}, Boss~\cite{Bos10}). Since sub-mm sized particles are well coupled to the gas, the same level of overdensity is expected for small grains as well. Considering that spiral waves are characterized by very extended structures, the results presented in this work show that the overdense regions in spiral waves can be optically thick even at mm-wavelengths and can even dominate the emission of a young disk at these wavelengths. This means that for gravitational unstable disks the measured low values of the mm-spectral index ($\alpha_{\rm{1-3mm}} \simless 3.0$) can be potentially explained by the optically thick emission of small grains from overdense regions in spiral waves. However, in order for disks to develope these instabilities, they need to be rather massive. Recent numerical simulations of Boss~(\cite{Bos10}) have shown that more than about 0.04 $M_{\odot}$ inside 20~AU are needed around a 1 $M_{\odot}$ young star, or about 0.02 $M_{\odot}$ around a 0.5 $M_{\odot}$ young star as considered in this work. By assuming a standard ISM-like value of 100 for the gas-to-dust ratio, this lower-limit corresponds to $2\cdot 10^{-4} M_{\odot}$ of dust mass inside 20~AU from the central protostar. Therefore, only the mm-fluxes of disks in the bright tail of the $F_{\rm{1mm}}-\alpha_{\rm{1-3mm}}$ diagram can be explained by accumulation of unprocessed small grains via gravitational instabilities. High sensitivity and angular resolution imaging with ALMA in the sub-mm will soon constrain the occurrence of these instabilities in real disks (Cossins et al.~\cite{Cos10}).

\section{Summary}
\label{sec:summary}

In this work we investigated the effect of possible local optically thick regions on the mm-wave emission of protoplanetary disks. The main goal of this exercise was to see what kind of structures are needed to explain the low values of spectral indices measured for protoplanetary disks in nearby star forming regions, without invoking the presence of mm-sized particles in the disk outer regions.
 
To quantify this potential effect we considered a simple disk model in which the disk is filled with regions with optically thick emission even at long millimeter wavelengths. In our calculations these regions occupy a fractional area of the disk which is left as a free parameter of the models, and we derive the predicted mm-fluxes for a range of different values for the parameters defining the structure of the unperturbed disk.

Interestingly, our analysis shows that relatively small filling factors for these optically thick regions can reproduce the measured fluxes of young disks in the (sub-)millimeter. In most cases, if the filling factor is kept constant throughout the disk, values lower than a few percent are sufficient to match the observed data. If the regions are instead localized in annuli with a width of $\sim 10$~AU, the required filling factors increase to a few $\times$~10\%. 

A significant local increase of the optical depth in the disk can be caused by the concentration of solid particles, as predicted by different physical mechanisms potentially acting in the disk.  
We discuss the physical plausibility of the required optically thick regions. In particular we investigate whether the required overdensities are reproduced in the simulations of the physical processes proposed in the literature to drive the concentration of particles. 

The main conclusion of this work is that for the vast majority of disks no physical processes proposed so far are capable to reproduce the measured low mm-spectral indices via a concentration of small ($<< 0.1$~mm) particles in optically thick regions. According to our current knowledge of these processes, only for the brightest disks, likely the most massive ones, the total mm-wave emission can be strongly affected by optically thick spiral arms and innermost dense regions. The results of this analysis further strengthen the scenario for which the measured low spectral indices of protoplanetary disks at long wavelengths are due to the presence of large mm/cm-sized pebbles in the disk outer regions.       
Observations of disks at (sub-)millimeter and centimeter wavelengths can therefore be used to constrain the models of the early phases of planetesimal formation.

\section*{Appendix A: New CARMA observations at 3mm}
\label{sec:obs}

We describe here new mm-wave observations of nine YSOs in the Taurus-Auriga star forming region (SFR). The sources were selected for
being relatively bright at sub-mm/mm wavelengths, namely with a flux density at 0.85~mm greater than 100~mJy and/or a 1.3 mm-flux density greater than 30~mJy. This criterion was chosen to have high chances to detect our sources at about 3~mm with the Combined Array for Research in Millimeter Astronomy (CARMA). The observed YSOs are listed in Table~1.

The dust thermal emission toward our sample of nine
young disk systems in Taurus-Auriga was observed with CARMA between 2010 March 2 and April 6. 
The array was in C configuration which provides baselines between
30 and 350 m. Observations were carried out at a central
frequency of 102.5~GHz (2.92~mm), with a total bandwidth
of 4~GHz.
The raw visibilities for each night were calibrated and
edited using the MIRIAD software package. Amplitude and
phase calibration were performed through observations of
the QSOs 0336+323 (for 04113+2758, V892 Tau, FN Tau), 3C 111 (for IC 2087, AB Aur, BP Tau, V836 Tau), 0449+113 (for DQ Tau). Passband calibration was obtained by
observing the QSO 3C 84. Mars and Uranus were used
to set the absolute flux scale. The uncertainty on CARMA
flux calibration is typically estimated to be ∼15\% and is
due to uncertainties in the planetary models and in the correction for atmospheric effects and hardware instabilities (see Ricci et al.~2011 for an analysis of the repeatability of 3mm-flux measurements with CARMA on observations performed in the same time period of the ones presented here).
Maps of the dust continuum emission were obtained by
standard Fourier inversion adopting natural weighting, and
photometry was obtained in the image plane. The resulting
FWHM of the synthesized beam is about 2$''$.

\begin{table*}
\centering \caption{Summary of the CARMA observations at 102.5~GHz. } \vskip 0.1cm
\begin{tabular}{lccrrc}
\hline
\hline

\\
Object  &     RA      &    DEC    & $F_{\nu}$ & rms   & Comments \vspace{1mm} \\
        &  (J2000)    &  (J2000)  &   (mJy)   & (mJy) &                        \\
  (1)   &     (2)     &    (3)    &    (4)    &  (5)  &     (6)
\\
\hline
\\
AB Aur        &  04:14:47.8     &   26:48:11.1 &  10.0    & 0.9 & Herbig Ae/Be \vspace{1mm} \\
V892 Tau      &  04:21:55.6     &   27:55:05.5 &  54.9    & 1.0 & Herbig Ae/Be \vspace{1mm} \\
BP Tau        &  04:30:44.3     &   26:01:23.9 &  10.1    & 0.6 & \vspace{1mm} \\
DQ Tau        &  04:38:28.6     &   26:10:49.7 &  14.9    & 0.6 & \vspace{1mm} \\
FN Tau        &  04:47:06.2     &   16:58:43.0 &   4.1    & 0.8 & extended envelope emission \vspace{1mm} \\
IC 2087$^{a}$ &  04:39:55.8     &   25:45:02.0 &  $< 2.7$ & 0.9 & Class I      \vspace{1mm} \\
V836 Tau      &  04:14:13.5     &   28:12:48.8 &  6.6     & 0.5 & \vspace{1mm} \\
MHO 1    	  &  04:14:17.0     &   28:10:56.5 &  50.5    & 1.1 & no sub-mm info \vspace{1mm} \\
MHO 2         &  04:14:47.8     &   26:48:11.1 &  30.2    & 1.1 & no sub-mm info \vspace{1mm} \\

\\
\hline
\end{tabular}

\begin{flushleft}
\textbf{Notes.} $^{(a)}$ For this undetected source the (RA,DEC) coordinates are from 2MASS (Cutri et al.~\cite{Cut03}). \\
Column~(6) reports the reason why the source has not been considered in the analysis (see Appendix A).
\end{flushleft}

\label{tab:sample}

\end{table*}

Among the nine observed YSOs, we considered the three sources (BP~Tau, DQ~Tau, V836~Tau) which satisfy the selection criteria adopted in this work and listed in Sect.~\ref{sec:sample}. 
We instead did not include the other observed sources for the following reasons:

\begin{itemize}
\item AB~Aur and V892~Tau, since they are Herbig Ae/Be stars, and therefore with a central PMS star more massive than the sample of T Tauri stars considered in this paper; 
\item FN~Tau shows evidence of spatially resolved nebulosity in near-infrared scattering light (Kudo et al.~\cite{Kud08}), likely due to a leftover protostellar envelope. Furthermore, Momose et al.~(\cite{Mom10}) showed how disk models fail to reproduce at the same time SMA 0.88~mm observations at high-angular resolution and single dish data at 1.3~mm, confirming that an extended ``halo'' component is present in the FN~Tau system.    
\item IC~2087, which is an embedded Class I YSO (Luhman et al.~\cite{Luh10}), and therefore, a significant contribution from an extended envelope to the sub-mm/mm emission is expected to occur;
\item MHO~1 and MHO~2, which are part of a binary system with angular separation of 4.0$''$ (or about 560~AU at the Taurus distance of 140~pc, Duchene et al.~\cite{Duc04}) and no interferometric observations which can separate the two components have been carried out so far in the sub-mm.
\end{itemize}

As for the three selected sources, according to Luhman et al.~(\cite{Luh10}), and references therein, BP~Tau is a K7-spectral type PMS star, DQ~Tau is a double lined spectroscopic binary which consists of two PMS stars with similar specctral type in the range of K7 to M1, and V836~Tau is a K7.
Andrews \& Williams~(\cite{And05}) reported fluxes for BP~Tau of 130$\pm$7~mJy and 47$\pm$0.7~mJy at about 0.85 and 1.3~mm, respectively, whereas for V836~Tau $F_{\rm{0.85mm}} = 74\pm3$~mJy, $F_{\rm{1.3mm}} = 37\pm6$~mJy. The other selected source, DQ~Tau, is a circumbinary disk which is known to undergo recurring millimeter flares due to star-star magnetic reconnection events (see Salter et al.~\cite{Sal10} and references therein). By measuring the light curve in the millimeter Salter et al.~(\cite{Sal10}) estimated a quiescent level of the emission at about 1.3~mm of 97~mJy, and of 17~mJy at 2.7~mm, which is roughly consistent with the measured flux of 14.9$\pm$0.6~mJy at 3.2~mm presented in this work. By combining these data with the new ones obtained at 3.2~mm with CARMA (Table~\ref{tab:sample}) for the three selected sources, we obtained the mm-spectral index shown as red empty circles in Fig.~\ref{fig:flux_alpha_fconst_vac30},\ref{fig:flux_alpha_fvar_gamma05_rc60}.

\section*{Appendix B: disk properties of the selected sample}

In Table~6 we report some of the properties of the 50 disks considered in this analysis. These properties were derived by fitting sub-mm data using two-layer disk models (for more details see the Ricci et al.~\cite{Ric10a},~\cite{Ric10a},~\cite{Ric10a},~\cite{Ric10a} papers). 

\begin{table*}[ht!]
\centering \vskip 0.1cm
\begin{footnotesize}
\begin{tabular}{lrcccc}
\hline \hline
Object name  & $F_{\rm{1mm}}$ & $\alpha_{\rm{1-3}mm}$  & $M_{\rm{dust}} \times
\kappa_{\rm{1 mm}}$ & $M_{\rm{dust}}^{q=2.5}$ & $M_{\rm{dust}}^{q=3}$   \vspace{1mm} \\
           & (mJy)  &              &  ($M_{\odot} \times$ cm$^2$g$^{-1}$) & ($M_{\odot}$) & ($M_{\odot}$) \\

\hline

\textit{Taurus} \vspace*{-1mm} \\ 

AA Tau  & 108 & 2.0 & $2.8\cdot10^{-4}$ & $1.5\cdot10^{-4}$ & $1.5\cdot10^{-3}$  \vspace*{-1mm} \\
BP Tau    & 85  & 1.9 & $1.9\cdot10^{-4}$ & $1.8\cdot10^{-4}$  & $5.0\cdot10^{-3}$   \vspace*{-1mm} \\
CI Tau  & 314 & 2.5 & $7.9\cdot10^{-4}$ & $1.1\cdot10^{-4}$ & $1.2\cdot10^{-4}$  \vspace*{-1mm} \\
CW Tau  & 129 & 2.5 & $3.2\cdot10^{-4}$ & $4.2\cdot10^{-5}$ & $4.4\cdot10^{-5}$  \vspace*{-1mm} \\
CX Tau  & 19  & 2.3 & $5.7\cdot10^{-5}$ & $1.2\cdot10^{-5}$ & $1.9\cdot10^{-5}$  \vspace*{-1mm} \\
CY Tau  & 168 & 2.2 & $6.0\cdot10^{-4}$ & $1.3\cdot10^{-4}$ & $2.0\cdot10^{-4}$  \vspace*{-1mm} \\
DE Tau  & 69  & 2.5 & $1.6\cdot10^{-4}$ & $2.4\cdot10^{-5}$ & $2.8\cdot10^{-5}$  \vspace*{-1mm} \\
DL Tau  & 313 & 2.4 & $8.4\cdot10^{-4}$ & $1.3\cdot10^{-4}$ & $1.5\cdot10^{-4}$  \vspace*{-1mm} \\
DM Tau  & 209 & 2.5 & $9.2\cdot10^{-4}$ & $1.0\cdot10^{-4}$ & $1.0\cdot10^{-4}$  \vspace*{-1mm} \\
DN Tau  & 153 & 2.3 & $3.5\cdot10^{-4}$ & $7.7\cdot10^{-5}$ & $1.2\cdot10^{-4}$  \vspace*{-1mm} \\
DO Tau  & 220 & 2.2 & $4.9\cdot10^{-4}$ & $1.4\cdot10^{-4}$ & $3.1\cdot10^{-4}$  \vspace*{-1mm} \\
DQ Tau   & 158  & 2.1 & $2.9\cdot10^{-4}$ & $1.9\cdot10^{-4}$  & $5.3\cdot10^{-3}$ \vspace*{-1mm} 
\\
DR Tau  & 298 & 2.7 & $6.6\cdot10^{-4}$ & $7.4\cdot10^{-5}$ & $7.2\cdot10^{-5}$  \vspace*{-1mm} \\
DS Tau  & 28  & 2.0 & $6.2\cdot10^{-5}$ & $4.0\cdot10^{-5}$ & $1.1\cdot10^{-3}$  \vspace*{-1mm} \\
FM Tau  & 29  & 1.9 & $6.5\cdot10^{-5}$ & $6.0\cdot10^{-5}$ & $1.7\cdot10^{-3}$  \vspace*{-1mm} \\
FZ Tau  & 27  & 2.1 & $5.0\cdot10^{-5}$ & $3.2\cdot10^{-5}$ & $9.0\cdot10^{-4}$  \vspace*{-1mm} \\
GM Aur  & 423 & 2.6 & $1.4\cdot10^{-3}$ & $1.6\cdot10^{-4}$ & $1.5\cdot10^{-4}$  \vspace*{-1mm} \\
GO Tau  & 151 & 2.6 & $6.3\cdot10^{-4}$ & $7.1\cdot10^{-5}$ & $6.9\cdot10^{-5}$  \vspace*{-1mm} \\
HO Tau  & 36  & 2.1 & $1.7\cdot10^{-4}$ & $3.7\cdot10^{-5}$ & $5.6\cdot10^{-5}$  \vspace*{-1mm} \\
IQ Tau  & 118 & 2.2 & $3.4\cdot10^{-4}$ & $7.4\cdot10^{-5}$ & $1.1\cdot10^{-4}$  \vspace*{-1mm} \\
RY Tau  & 383 & 2.3 & $5.1\cdot10^{-4}$ & $1.1\cdot10^{-4}$ & $1.7\cdot10^{-4}$  \vspace*{-1mm} \\
SU Aur  & 50  & 2.1 & $5.0\cdot10^{-5}$ & $3.2\cdot10^{-5}$ & $9.0\cdot10^{-4}$  \vspace*{-1mm} \\
UZ Tau E & 333 & 2.5 & $6.1\cdot10^{-4}$ & $9.6\cdot10^{-5}$ & $1.1\cdot10^{-4}$ \vspace*{-1mm} 
\\
V836 Tau  & 57  & 1.9 & $1.3\cdot10^{-4}$ & $1.2\cdot10^{-4}$  & $3.4\cdot10^{-3}$  \vspace*{-1mm} \\
\hline 
\textit{Ophiuchus} \vspace*{-1mm} \\

SR 4    & 79  & 2.5 & $1.3\cdot10^{-4}$ & $2.1\cdot10^{-5}$ & $2.4\cdot10^{-5}$  \vspace*{-1mm} \\
GSS 26  & 215 & 1.9 & $3.5\cdot10^{-4}$ & $2.3\cdot10^{-3}$ &  ...               \vspace*{-1mm} \\
EL 20   & 151 & 2.5 & $3.2\cdot10^{-4}$ & $4.3\cdot10^{-5}$ & $4.5\cdot10^{-5}$  \vspace*{-1mm} \\
DoAr 25 & 405 & 2.3 & $8.0\cdot10^{-4}$ & $1.8\cdot10^{-4}$ & $2.6\cdot10^{-4}$  \vspace*{-1mm} \\
EL 24   & 664 & 2.2 & $9.9\cdot10^{-4}$ & $2.9\cdot10^{-4}$ & $6.3\cdot10^{-4}$  \vspace*{-1mm} \\
EL 27   & 564 & 2.2 & $1.5\cdot10^{-3}$ & $3.5\cdot10^{-4}$ & $5.6\cdot10^{-4}$  \vspace*{-1mm} \\
SR 21   & 220 & 2.9 & $5.3\cdot10^{-4}$ & $4.9\cdot10^{-5}$ & $4.5\cdot10^{-5}$  \vspace*{-1mm} \\
IRS 41  & 84  & 2.1 & $1.3\cdot10^{-4}$ & $6.6\cdot10^{-5}$ & $6.8\cdot10^{-4}$  \vspace*{-1mm} \\
YLW 16c & 123 & 2.4 & $2.3\cdot10^{-4}$ & $4.3\cdot10^{-5}$ & $5.6\cdot10^{-5}$  \vspace*{-1mm} \\
IRS 49  & 40  & 1.8 & $3.9\cdot10^{-5}$ & $2.2\cdot10^{-3}$ & ...                \vspace*{-1mm} \\
DoAr 33 & 64  & 2.2 & $1.2\cdot10^{-4}$ & $3.4\cdot10^{-5}$ & $9.9\cdot10^{-5}$  \vspace*{-1mm} \\
WSB 52  & 88  & 1.8 & $1.4\cdot10^{-5}$ & $2.6\cdot10^{-3}$ & ...                \vspace*{-1mm} \\
WSB 60  & 156 & 1.9 & $5.6\cdot10^{-4}$ & $2.9\cdot10^{-4}$ & $3.0\cdot10^{-3}$  \vspace*{-1mm} \\
DoAr 44 & 168 & 2.2 & $3.0\cdot10^{-4}$ & $8.8\cdot10^{-5}$ & $1.9\cdot10^{-4}$  \vspace*{-1mm} \\
RNO 90  & 111 & 2.3 & $1.1\cdot10^{-4}$ & $3.1\cdot10^{-5}$ & $7.1\cdot10^{-5}$  \vspace*{-1mm} \\
Wa Oph 6 & 250 & 2.4 & $4.9\cdot10^{-4}$ & $8.0\cdot10^{-5}$ & $9.8\cdot10^{-5}$ \vspace*{-1mm} \\
AS 209  & 441 & 2.4 & $7.9\cdot10^{-4}$ & $1.2\cdot10^{-4}$ & $1.4\cdot10^{-4}$  \vspace*{-1mm} \\
\hline
\textit{Orion Nebula Cluster} \vspace*{-1mm} \\

121-1925  & 14 & 1.8 & $3\cdot10^{-4}$ & $1\cdot10^{-3}$ & ... \vspace*{-1mm} \\
136-1955  & 64 & 3.4 & $1\cdot10^{-4}$ & $6\cdot10^{-5}$ & $6\cdot10^{-5}$ \vspace*{-1mm} \\
141-1952  & 24 & 2.3 & $6\cdot10^{-5}$ & $3\cdot10^{-5}$ & $5\cdot10^{-5}$ \vspace*{-1mm} \\
181-825   & 44 & 2.0 & $5\cdot10^{-4}$ & $5\cdot10^{-4}$ & $6\cdot10^{-3}$ \vspace*{-1mm} \\
216-0939  & 64 & 2.3 & $8\cdot10^{-4}$ & $2\cdot10^{-4}$ & $3\cdot10^{-4}$ \vspace*{-1mm} \\
253-1536a & 104 & 2.2 & $1\cdot10^{-3}$ & $7\cdot10^{-4}$ & $1\cdot10^{-3}$ \vspace*{-1mm} \\
253-1536b & 28 & 1.9 & $5\cdot10^{-3}$ & $6\cdot10^{-4}$ & $8\cdot10^{-3}$ \vspace*{-1mm} \\
132-1832  & 16 & $>$1.5 & ...            & ...               & ...            \vspace*{-1mm} \\
280-1720  & 21 & $>$2.1 & ...            & ...               & ...            \vspace*{-1mm} \\

\hline
\end{tabular}

\end{footnotesize}

\vspace*{-3mm}

\begin{flushleft}

\caption{Columns (2) and (3) report the source flux density at 1.0 mm and spectral index between 1 and 3~mm, respectively, from the best fit two-layer disk models adopted in Ricci et al.~(\cite{Ric10a}, \cite{Ric10b}, \cite{Ric11a}, \cite{Ric11b}). The uncertainties of fluxes in the millimeter range are typically dominated by the uncertainty on the absolute flux scale of the observations. Values of these uncertainties depend on the particular instrument and conditions during the observations, but have typical values around $10-20\%$. Contrary to Figures 1 and 2, the 1~mm-fluxes reported here do not consider any correction for the different distances of the three regions. Column (4) lists the product between dust mass and dust opacity at 1~mm from the same models. Columns (5) and (6) report the inferred dust mass for values of 2.5 and 3, respectively, for the power-law slope $q$ of the grain size numberdensity (see e.g. Ricci et al.~\cite{Ric10a} for more details). Blank fields in Columns (4), (5), (6) are for cases in which no good fit could be obtained.}

\end{flushleft}

\label{tab:fits_results}
%\begin{flushleft}
%$a$) References - A90: \cite{Ada90}; B91: \cite{Bec91}; M94: \cite{Man94}; A94: \cite{Alt94}; D96: \cite{Dut96}; M96: \cite{Mun96}; K02: \cite{Kit02}; A05: \cite{And05}; R06: \cite{Rod06}; A07: \cite{And07}; I09: \cite{Ise09}.
%\end{flushleft}
\end{table*}

%%%%%%%%%%%%%%%%%%%%%%%% acknowledgments
\begin{acknowledgements}
L.R. aknowledges the PhD fellowship of the International Max-Planck-Research School. 
L.T. and F.T. acknowledge support from ASI under contract with INAF-Osservatorio Astrofisico di Arcetri.
Support for CARMA construction was derived from the states of California,
Illinois, and Maryland, the James S. McDonnell Foundation, the
Gordon and Betty Moore Foundation, the Kenneth T. and Eileen
L. Norris Foundation, the University of Chicago, the Associates of
the California Institute of Technology, and the National Science
Foundation. Ongoing CARMA development and operations are sup-
ported by the National Science Foundation under a cooperative
agreement, and by the CARMA partner universities.
\end{acknowledgements}
%%%%%%%%%%%%%%%%%%%%%%%%%%%%%%%%%%%%%

%%%%%%%%%%%%%% references %%%%%%%%%%%%%%%%

%%%%%%%%%%%%%%%%%%%%%%%%%%%%%%


\begin{thebibliography}{}
%\bibitem[Adams et al. (1990)]{Ada90} Adams, F. C., Emerson, J. P., \&
%Fuller, G. A. 1990,
%ApJ, 357, 606
%\bibitem[Altenhoff et al. (1994)]{Alt94} Altenhoff, W. J., Thum, C., \&
%Wendker, H. J. 1994,
%A\&A, 281, 161
\bibitem[2011]{And11} Andrews, S. M., Wilner, D. J., Espaillat, C., Hughes, A. M., Dullemond, C. P., McClure, M. K., Qi, C., \& Brown, J. M. 2011, ApJ 732, 42
\bibitem[2010]{And10} Andrews, S. M., Wilner, D. J., Hughes, A. M., Qi, C., \& Dullemond, C. P. 2010, ApJ 723, 1241
\bibitem[2009]{And09} Andrews, S. M., Wilner, D. J., Hughes, A. M., Qi, C., \& Dullemond, C. P. 2009, ApJ 700, 1502
\bibitem[2005]{And05} Andrews, S. M., \& Williams, J. P. 2005, ApJ 631, 1134
%\bibitem[2007a]{And07a} Andrews, S. M., \& Williams, J. P. 2007, ApJ 671, 1800
%\bibitem[2007b]{And07b} Andrews, S. M., \& Williams, J. P. 2007, ApJ 659, 705
%\bibitem[Baraffe et al. (1998)]{Bar98} Baraffe, I., Chabrier, G., Allard, F. \&
%Hauschildt, P. H. 1998,
%A\&A, 337, 403
\bibitem[1990]{Bec90} Beckwith, S. V. W., Sargent, A. I., Chini, R. S., \& Guesten, R. 1990,
%AJ, 99, 924
\bibitem[1991]{Bec91} Beckwith, S. V. W., \& Sargent, A. I. 1991, ApJ 381, 250
%\bibitem[2000]{Bec00} Beckwith, S. V. W., Henning, Th., \& Nakagawa, Y., in Mannings, V., Boss, A. P., Russell, S. S. (eds.), Protostar \& Planets IV. Univ. of Arizona Press. Tucson 2000. p. 533 
\bibitem[1999]{Ber99} Bertout, C., Robichon, N., \& Arenou, F.
%1999,
%A\&A, 352, 574
%\bibitem[Bessell \& Brett (1988)]{Bes88} Bessell, M. S., \& Brett J. M. 1988, PASP 100, 1134
\bibitem[2010a]{Bir10a} Birnstiel, T., Ricci, L., Trotta, F., Dullemond, C. P., Natta, A., Testi, L., Dominik, C., Henning, T., Ormel, C. W., \& Zsom, A. 2010, A\&A 516L, 14
%\bibitem[2010b]{Bir10b} Birnstiel, T., Dullemond, C. P., \& Brauer, F. 2010, A\&A 513, 79
%\bibitem[2009]{Bir09} Birnstiel, T., Dullemond, C. P., \& Brauer, F. 2009, A\&A 503, 5
\bibitem[2010]{Bos10} Boss, A. P. 2010, ApJ 725, 145
%\bibitem[2001]{Bou01} Bouwman, J., Meeus, G., de Koter, A., Hony, S., Dominik, C., \& Waters, L. B. F. M. 2001, A\&A 375, 950
\bibitem[2007]{Bra07} Brauer, F., Dullemond, C. P., Johansen, A., et al. 2007, A\&A 469, 1169 
%\bibitem[Brice\~no et al. (1998)]{Bri98} Brice\~no, C., Hartmann, L., Stauffer, J. R., \& Martin, E. L.
%1998, AJ, 115, 2074
%\bibitem[Brice\~no et al. (2002)]{Bri02} Brice\~no, C., Luhman, K. L., Hartmann L., Stauffer, J. R., \& Kirkpatrick, J. D. 2002, ApJ 580, 317
%\bibitem[2009]{Bro09} Brown, J. M., Blake, G. A., Qi, C., Dullemond, C. P., Wilner, D. J., \& Williams, J. P. 2009, ApJ 704, 496
%\bibitem[2001]{Bou01} Bouwman, J., Meeus, G., de Koter, A., Hony, S., Dominik, C., \& Waters, L. B. F. M. 2001, A\&A 375, 950
%\bibitem[1989]{Car89} Cardelli, J. A., Clayton G. C., \& Mathis, J. S. 1989, ApJ 345, 245
%\bibitem[Carpenter (2001)]{Car01} Carpenter, J. M. 2001, AJ 121, 2851
\bibitem[2010]{Chi10} Chiang, E., \& Youdin, A. N. 2010, AREPS 38, 493
\bibitem[1997]{Chi97} Chiang, E., \& Goldreich, P. 1997, ApJ 490, 368
\bibitem[2010]{Cos10} Cossins, P., Lodato, G., \& Testi, L. 2010, MNRAS 407, 181
\bibitem[2003]{Cut03} Cutri, R. M. et al. 2003, 2MASS All Sky Catalog of point sources
%\bibitem[1998]{Den98} Dent, W. R. F., Matthews, H. E., \& Ward-Thompson, D. 1998, MNRAS 301, 1049
\bibitem[2006]{Dra06} Draine, B. T. 2006, ApJ 636, 1114
%\bibitem[2007]{Dom07} Dominik, C., Blum, J., Cuzzi, J. N., \& Wurm, G., in Reipurth, B., Jewitt, D., Keil, K. (eds.), Protostars \& Planets V. University of Arizona Press. Tucson. 2007. p. 783
\bibitem[2004]{Duc04} Duchene, G., Bouvier, J., Bontemps, S., Andr\'e, P., Motte, F. 2004, A\&A 427, 651
\bibitem[2001]{Dul01} Dullemond, C. P., Dominik, C., \& Natta, A. 2001, ApJ, 560, 957
%\bibitem[Dutrey et al. (1996)]{Dut96} Dutrey, A., Guilloteau, S., Duvert, G., et al. 1996,
%A\&A, 309, 493
\bibitem[2005]{Fro05} Fromang, S., \& Nelson, R. P. 2005, MNRAS 364, 81L
%\bibitem[2006]{Fur06} Furlan, E., Hartmann, L., Calvet, N., et al. 2006, ApJS 165, 568
\bibitem[2011]{Gui11} Guilloteau, S., Dutrey, A., Pietu, V., \& Boehler, Y. 2011, A\&A 529, 105
%\bibitem[1998]{Gul98} Gullbring, E., Hartmann, L., Brice\~no, C., \& Calvet, N. 1998, ApJ 492, 323
%\bibitem[Hartigan 
%\& Kenyon (2003)]{Har03} Hartigan, P., \& Kenyon, S.~J. 2003, ApJ, 583, 334 
%\bibitem[Herczeg \& Hillenbrand (2008)]{Her08} Herczeg, C. J., \& Hillenbrand L. A. 2008, ApJ 681, 594
\bibitem[2010]{Ise10} Isella, A., Carpenter, J. M., \& Sargent, A. I. 2010, ApJ 714, 1746
\bibitem[2009]{Ise09} Isella, A., Carpenter, J. M., \& Sargent, A. I. 2009, ApJ 701, 260
%\bibitem[Itoh et al. (2005)]{Ito05} Itoh, Y., et al. 2009, ApJ 620, 984
\bibitem[2009]{Joh09} Johansen, A., Youdin, A., \& Klahr, H. 2009, ApJ 697, 1269
\bibitem[2007]{Joh07} Johansen, A., \& Youdin, A. 2007, ApJ 662, 627
%\bibitem[2007]{Jor07} J\o rgensen, J. K., Bourke, T. L., Myers, P. C., Di Francesco, J., van Dishoeck, E. F., Lee, Chin-Fei, Ohashi, N., Schoeier, F. L., Takakuwa, S., Wilner, D. J., \& Zhang, Q. 2007, ApJ 659, 479
%\bibitem[Kenyon \& Hartmann (1995)]{Ken95} Kenyon, S. J., \& Hartmann L. 1995, ApJS 101, 117
%\bibitem[2006]{Kes06} Kessler-Silacci, J.; Augereau, J. C.; Dullemond, C. P., Geers, V., Lahuis, F., et al. 2006, ApJ 639, 275
%\bibitem[Kitamura et al. (2002)]{Kit02} Kitamura, Y., Momose, M., Yokogawa, S., et al. 2002, ApJ 581, 357
\bibitem[2003]{Kla03} Klahr, H. H., \& Bodenheimer, P. 2003, ApJ 582, 869
%\bibitem[Kr\"ugel \& Siebenmorgen (1994)]{Kru94} Kr\"ugel, E., \& Siebenmorgen, R. 1994, A\&A 288, 929
%\bibitem[2009]{Kwo09} Kwon, W., Looney, L. W., Mundy, L. G., Chiang, H.-F., \& Kemball, A. J. 2009, ApJ 696, 841
\bibitem[2008]{Kud08} Kudo, T., et al. 2008, ApJL 673, 67
%\bibitem[Leggett (1992)]{Leg92} Leggett, S. K. 1992, ApJS 82, 351
%\bibitem[Leinert et 
%al. (1993)]{Lei93} Leinert, C., Zinnecker, H., Weitzel, N., Christou, J., Ridgway, S.~T., Jameson, R., Haas, M., \& Lenzen, R. 1993, A\&A, 278, 129 
%\bibitem[2008]{Loi08} Loinard, L., Torres, R. M., Mioduszewski, Ami, J., \& Rodriguez, L. F. 2008, ApJ 675, 29
%\bibitem[2008]{Lom08} Lombardi, M., Lada, C. J., \& Alves, J. 2008, A\&A 480, 785
%\bibitem[2007]{Lom07} Lommen, D., Wright, C. M., Maddison, S. T., Jørgensen, J. K., Bourke, T. L., van Dishoeck, E. F., Hughes, A., Wilner, D. J., Burton, M., \& van Langevelde, H. J. 2007, A\&A 462, 211
%\bibitem[2010]{Lom10} Lommen, D. J. P., van Dishoeck, E. F., Wright, C. M., Maddison, S. T., Min, M.; Wilner, D. J., Salter, D. M., van Langevelde, H. J., Bourke, T. L., van der Burg, R. F. J., \& Blake, G. A. 2010, arXiv1004:0158
\bibitem[1999]{Lov99} Lovelace, R. V. E., Li, H., Colgate, S. A., \& Nelson, A. F. 1999, ApJ 513, 805
\bibitem[2010]{Luh10} Luhman, K. L., Allen, P. R., Espaillat, C., Hartmann, L., Calvet, N. 2010, ApJS 186, 111
%\bibitem[1999]{Luh99} Luhman, K. L. 1999, ApJ 525, 466
\bibitem[1974]{Lyn74} Lynden-Bell, D., \& Pringle, J. E. 1974, MNRAS, 168, 603
\bibitem[2009]{Lyr09} Lyra, W., Johansen, A., Zsom, A., Klahr, H., \& Piskunov, N. 2009, A\&A 497, 869
%\bibitem[Mannings \& Emerson (1994)]{Man94} Mannings, V., \& Emerson, J. P. 1994, MNRAS, 267, 361
\bibitem[1977]{Mat77} Mathis, J. S., Rumpl, W., \& Nordsieck, K. H. 1977, ApJ 217, 425
%\bibitem[2006]{McC06} McCabe, C., Ghez, A. M., Prato, L., Duchene, G., Fisher, R. S., \& Telesco, C. 2006, ApJ 636, 932
%\bibitem[2010]{McC10} McClure, M. K., Furlan, E., Manoj, P., Luhman, K. L., Watson, D. M., Forrest, W. J., Espaillat, C., Calvet, N., D'Alessio, P., Sargent, B., Tobin, J. J., \& Chiang, H.-F. 2010, ApJS 188, 75 
\bibitem[2007]{Men07} Menten, K. M., Reid, M., et al. 2007, A\&A 474, 515
%\bibitem[1997]{Mey97} Meyer, M. R., Calvet, N., \& Hillenbrand, L. A. 1997, AJ 114, 288
%\bibitem[Miyake \& Nakagawa (1993)]{Miy93} Miyake, K., \& Nakagawa, Y. 1993, Icarus, 106, 20
\bibitem[2010]{Mom10} Momose, M., Ohashi, N., Kudo, T., Tamura, M., Kitamura, Y. 2010, ApJ 712, 397
%\bibitem[1998]{Mot98} Motte, F., Andre, P., \& Neri, R. 1998, A\&A 336, 150
%\bibitem[Mundy et al. (1996)]{Mun96} Mundy, L. G., et al. 1996, ApJ 464, L169
\bibitem[2007]{Nat07} Natta, A., Testi, L., Calvet, N., Henning, Th., Waters, R., \& Wilner, D., in Reipurth, B., Jewitt, D., Keil, K. (eds.), Protostars \& Planets V. University of Arizona Press. Tucson. 2007. p. 783
%\bibitem[2006]{Nat06} Natta, A., Testi, L., \& Randich, S. 2006, A\&A 452, 245
\bibitem[2004]{Nat04} Natta, A., Testi, L., Neri, R., Shepherd, D. S., \& Wilner, D. J. 2004, A\&A 416, 179
%\bibitem[2009]{Orm09} Ormel, C. W., Paszun, D., Dominik, C., \& Tielens, A. G. G. M. 2009, A\&A 502, 845
%\bibitem[1999]{Pal99} Palla, F., \& Stahler, S. W. 1999, ApJ 525, 772
%\bibitem[Pinte et 
%al.(2007)]{Pin07} Pinte, C., Fouchet, L., M{\'e}nard, F., Gonzalez, J.-F., \& Duch{\^e}ne, G. 2007, A\&A, 469, 963 
%\bibitem[1994]{Pol94} Pollack, J. B., Hollenbach, D., Beckwith, S., Simonelli, D. P., Roush, T., \& Fong, W. 1994, ApJ 421, 615
%\bibitem[2005]{Rat05} Ratzka, T., K\"ohler, R., \& Leinert, Ch. 2005, A\&A 437, 611
\bibitem[2011]{Reg11} Regaly, Zs., Juhasz, A., Sandor, Zs., \& Dullemond, C. P. 2011, arXiv:1109.6177
%\bibitem[1993]{Rei93} Reipurth, B., \& Zinnecker, H. 1993, A\&A 278, 81
%\bibitem[Rieke \& Lebofsky (1985)]{Rie85} Rieke, G. H., \&
%Lebofsky, M. J. 1985, ApJ 288, 618
\bibitem[2011b]{Ric11b} Ricci, L., Testi, L., Williams, J. P., Mann, R. K., \& Birnstiel, T. 2011b, ApJ Letters 739, 8
\bibitem[2011a]{Ric11a} Ricci, L., Mann, R. K., Testi, L., Williams, J. P., Isella, A., Robberto, M., Natta, A., \& Brooks, K. 2011a, A\&A 525, 81
\bibitem[2010b]{Ric10b} Ricci, L., Testi, L., Natta, A., \& Brooks, K. 2010b, A\&A 521, 66
\bibitem[2010a]{Ric10a} Ricci, L., Testi, L., Natta, A., Neri, R., Cabrit, S., \& Herczeg. G. J. 2010a, A\&A 512, 15
\bibitem[2004]{Ric04} Rice, W. K. M., Lodato, G., Pringle, J. E., Armitage, P. J., \& Bonnell, I. A. 2004, MNRAS 355, 543
\bibitem[2006]{Rod06} Rodmann, J., Henning, T., Chandler, C. J., Mundy, L. G., \& Wilner, D. J. 2006, A\&A 446, 211
\bibitem[2010]{Sal10} Salter, D. M., et al. 2010, A\&A 521, 32
%\bibitem[1982]{Sch82} Schmidt-Kaler T. 1982, in Landolt-BornsteinGroup VI, Vol. 2, ed. K.-H. Hellwege (Berlin: Springer), 454
\bibitem[2003]{Sem03} Semenov, D., Henning, T., Helling, C., Ilgner, M., \& Sedlmayr, E. 2003, A\&A, 410, 611 
%\bibitem[1995]{Sim95} Simon, M., Ghez, A. M., Leinert, Ch., Cassar, L., Chen, W. P., Howell, R. R., Jameson, R. F., Matthews, K., Neugebauer, G., \& Richichi, A. 1995, ApJ, 443, 625 
\bibitem[2001]{Tes01} Testi, L., Natta, A., Shepherd, D. S., \& Wilner, D. J. 2001, ApJ 554, 1087
\bibitem[2003]{Tes03} Testi, L., Natta, A., Shepherd, D. S., \& Wilner, D. J. 2003, A\&A 403, 323
%\bibitem[2003]{Van03} Van Boekel, R., Waters, L. B. F. M., Dominik, C., Bouwman, J., de Koter, A., Dullemond, C. P., Paresce, F. 2003, A\&A 400, 21
\bibitem[1984]{War84} Warren, S. G. 1984, ApOpt 23, 1206
\bibitem[1977]{Wei77} Weidenschilling, S. J. 1977, MNRAS 180, 57
\bibitem[2001]{Wei01} Weingartner, J. C., \& Draine, B. T. 2001, ApJ 548, 296
\bibitem[2008]{Wil08} Wilking, B. A., Gagn\'e, M., \& Allen, L. E. 2008, in Handbook of Star Forming Regions: Volume II, The Southern Sky, ed. Reipurth (San Francisco, CA: ASP), 351
\bibitem[2000]{Wil00} Wilner, D. J., Ho, P. T. P., Kastner, J. H., \& Rodriguez, L. F. 2000, ApJL 534, 101
\bibitem[2005]{Wil05} Wilner, D. J., D'Alessio, P., Calvet, N., et al. 2005, ApJ Letters 626, 109
\bibitem[1996]{Zub96} Zubko, V. G., Mennella, V., Colangeli, L., \& Bussoletti, E. 1996, MNRAS, 282, 1321


\end{thebibliography}
\end{document}